\documentclass[envcountsame,orivec]{llncs}
\usepackage{proof}
\usepackage{amssymb}
\usepackage{fullpage}
\usepackage{graphics} 

\newcommand{\NEW}{\reflectbox{\ensuremath{\mathsf{N}}}}

\def\relbar{\mathrel{\smash-}}
\def\joinrelm{\mathrel{\mkern-3mu}}
\def\tailpiece{\kern 1pt\vrule height 1ex width 0.3ex depth -.3ex}
\def\seqsym{\mathrel{\tailpiece\joinrelm\relbar}}
\newcommand{\Seq}[2]{#1\seqsym #2}
\newcommand{\NSeq}[3]{#1 ; #2 \seqsym #3}
\newcommand{\TSeq}[3]{#1 \vdash #2 : #3}

\newcommand{\botL}{\bot{\cal L}}
\newcommand{\cL}{\hbox{\sl c}{\cal L}}
\newcommand{\defeq}{\stackrel{{\scriptstyle\triangle}}{=}}
\newcommand{\defL}{\hbox{\sl def}{\cal L}}
\newcommand{\defR}{\hbox{\sl def\/}{\cal R}}

\newcommand{\eqL}{{\rm eq}{\cal L}}
\newcommand{\eqR}{{\rm eq}{\cal R}}
\newcommand{\existsL}{\exists{\cal L}}
\newcommand{\existsR}{\exists{\cal R}}
\newcommand{\forallL}{\forall{\cal L}}
\newcommand{\forallR}{\forall{\cal R}}
\newcommand{\landL}{\land{\cal L}}
\newcommand{\landR}{\land{\cal R}}
\newcommand{\lorL}{\lor{\cal L}}
\newcommand{\lorR}{\lor{\cal R}}
\newcommand{\nablaL}{\nabla{\cal L}}
\newcommand{\nablaR}{\nabla{\cal R}}
\newcommand{\oimp}{\supset}
\newcommand{\oimpL}{\oimp{\cal L}}
\newcommand{\oimpR}{\oimp{\cal R}}

\newcommand{\topR}{\top{\cal R}}
\newcommand{\wL}{\hbox{\sl w}{\cal L}}

\newcommand{\FOL   }{FO\lambda}
\newcommand{\Nset}{{\rm I} \! {\rm N}}
\newcommand{\FOLDN }{\FOL^{\Delta\Nset}}
\newcommand{\LGN}{LG^\omega}
\newcommand{\LG}{LG}

\newcommand{\oforall}{\bigwedge}

\newcommand{\oand}{\mathbin{\&}}

\newcommand{\FOLNb }{\FOL^{\nabla}}
\newcommand{\FOLDNb}{FO\lambda^{\Delta\nabla}}

\newlength{\infwidthi}
\newlength{\infwidthii}
\newcommand{\init}{id}
\newcommand{\mc}{mc}
\newcommand{\natL}{nat{\cal L}}
\newcommand{\natR}{nat{\cal R}}

\begin{document}

\title{Cut Elimination for a Logic with Generic Judgments and Induction}
\author{Alwen Tiu}
\institute{Computer Sciences Laboratory\\ 
The Australian National University}

\maketitle

\begin{abstract}
This paper presents a cut-elimination proof for the logic $LG^\omega$, 
which is an extension of a proof system for encoding generic judgments,
the logic $\FOLDNb$ of Miller and Tiu, with an induction principle.
The logic $LG^\omega$, just as $\FOLDNb$, features extensions of first-order
intuitionistic logic with fixed points and a ``generic quantifier'', $\nabla$, 
which is used to reason about the dynamics of bindings in object systems encoded 
in the logic. A previous attempt to extend $\FOLDNb$ with an 
induction principle has been unsuccessful in modeling some
behaviours of bindings in inductive specifications. 
It turns out that this problem can be solved by relaxing 
some restrictions on $\nabla$, 
in particular by adding the axiom $B \equiv \nabla x. B$, where
$x$ is not free in $B$. We show that by adopting the
equivariance principle, the presentation of the extended logic
can be much simplified. 
This paper contains the technical proofs for the 
results stated in \cite{tiu07entcs}; readers are encouraged to consult \cite{tiu07entcs} for
motivations and examples for $LG^\omega.$
\end{abstract}

\section{Introduction}

This work aims at providing a framework for reasoning about
specifications of deductive systems using {\em higher-order
abstract syntax}~\cite{pfenning88pldi}. 
Higher-order abstract syntax is a declarative approach to encoding syntax
with bindings using Church's simply typed $\lambda$-calculus. 
The main idea is to support the notions of $\alpha$-equivalence 
and substitutions in the object syntax by operations
in $\lambda$-calculus, in particular $\alpha$-conversion and
$\beta$-reduction. There are at least two approaches to
higher-order abstract syntax. The {\em functional programming} approach
encodes the object syntax as a data type, where the binding constructs
in the object language are mapped to functions in the functional language.
In this approach, terms in the object language become values of
their corresponding types in the  functional language. 
The {\em proof search} approach encodes object syntax as 
expressions in a logic whose terms are simply typed, 
and functions that act on the object terms are defined 
via relations, i.e., logic programs. 
There is a subtle difference between
this approach and the former; 
in the proof search approach,
the simple types are inhabited by well-formed expressions, instead
of values as in the functional approach (i.e., the abstraction type
is inhabited by functions). 
The proof search approach is often referred to 
as {\em $\lambda$-tree syntax}~\cite{miller99surveys}, to distinguish 
it from the functional approach. This paper concerns the $\lambda$-tree syntax
approach.

Specifications which use $\lambda$-tree syntax are often
formalized using hypothetical and generic judgments
in intuitionistic logic. It is enough to restrict to
the fragment of first-order intuitionistic logic whose only
formulas are those of hereditary Harrop formulas, which we will
refer to as the $HH$ logic. 
Consider for instance the problem of 
defining the data type for untyped $\lambda$-terms. 
One first introduces the following constants:
$$
app : tm \to tm \to tm
\qquad
abs : (tm \to tm) \to tm
$$
where the type $tm$ denotes the syntactic category of $\lambda$-terms
and $app$ and $abs$ encode application and abstraction, respectively.
The property of being a $\lambda$-term is then defined via 
the following theory:
$$
\oforall M\oforall N(lam~M \land lam~N \Rightarrow lam~(app~M~N)) ~\oand~
$$
$$
\oforall M ( (\oforall x. lam~x \Rightarrow lam~(M\,x)) \Rightarrow 
lam~(abs~M))
$$
where $\oforall$ is the universal quantifier and $\Rightarrow$ is
implication. 

Reasoning about object systems encoded in $HH$ is reduced to reasoning
about the structure of proofs in $HH$. McDowell and Miller
formalize this kind of reasoning in the logic $\FOLDN$~\cite{mcdowell00tcs}, 
which is an extension of first-order intuitionistic logic with fixed points
and natural numbers induction. This is done by 
encoding the sequent calculus of $HH$ inside $\FOLDN$ and prove
properties about it. We refer to $HH$ as object logic and 
$\FOLDN$ as meta logic.
McDowell and Miller considered different styles of encodings and concluded 
that explicit representations of hypotheses and, more importantly,
eigenvariables of the object logic are required in order to capture some
statements about object logic provability in the meta 
logic~\cite{mcdowell02tocl}. 
One typical example involves the use of hypothetical and generic reasoning
as follows: Suppose that the following formula is provable in $HH$.
$$
\oforall x. p\,x\,s \Rightarrow \oforall y. p\,y\,t \Rightarrow p\,x\,t.
$$
By inspection on the inference rules of $HH$, one observes 
that this is only possible if $s$ and $t$ are syntactically equal.
This observation comes from the fact that the right introduction
rule for universal quantifier, 
reading the rule bottom-up, introduces new constants, 
or eigenvariables.
The quantified variables $x$ and $y$ will be replaced by
distinct eigenvariables and hence the only matching hypothesis
for $p\,x\,t$ would be $p\,x\,s$, and therefore $s$ and $t$
has to be equal.  
Let $\vdash_{HH} F$ denote the provability of the formula $F$ in $HH$.
Then in the meta logic, we would want to be able to prove the statement:
$$
\forall s\forall t. (\vdash_{HH} \oforall x. p\,x\,s \Rightarrow \oforall y. p\,y\,t \Rightarrow p\,x\,t) \oimp s = t.
$$
The question is then how we would intrepret the object logic eigenvariables
in the meta logic. It is demonstrated in ~\cite{mcdowell02tocl} 
that the existing quantifiers in $\FOLDN$ cannot be used to capture 
the behaviours
of object logic eigenvariables directly. McDowell and Miller then 
resort to a non-logical encoding technique (in the sense that no
logical connectives are used) which has some similar flavor to the use
of deBruijn indices. The use of this encoding technique, however, 
has a consequence that substitutions in the object logic has to be 
formalized explicitly. 

Motivated by the above mentioned limitation of $\FOLDN$,
Miller and Tiu later introduce a new quantifier $\nabla$ to $\FOLDN$
which allows one to move the binders from the object logic
to the meta logic. A generic judgment in the object logic,
for instance $\vdash_{HH} \oforall x.G\,x$ is reflected
in the meta logic as $\nabla x.\vdash_{HH} G\,x.$
This meta logic, called $\FOLDNb$~\cite{miller05tocl}, 
allows one to perform case
analyses on the provability of the object logic. 
Tiu later extended $\FOLDNb$ with induction and co-induction
rules, resulting in the logic Linc~\cite{tiu04phd}. 
However, some inductive properties about the object logic
are not provable in Linc. For example, the fact that 
$\vdash_{HH} \oforall x.G\,x$ implies
$\forall t. \vdash_{HH} G\,t$ (that is, the extensional
property of universal quantification) is not provable
in Linc. As it is shown in \cite{tiu04phd}, this is partly 
caused by the fact that $B \equiv \nabla x.B$, where 
$x$ is not free in $B$, is not provable in Linc or $\FOLDNb$. 
In this paper we present the logic $\LGN$, which is an
extension of $\FOLDNb$ with natural number induction and
with the axiom schemes:
\begin{equation}
\label{eq1}
\nabla x\nabla y.B \, x\,y \oimp \nabla y\nabla x.B\,x\,y
\quad \hbox{ and } \quad 
B \equiv \nabla x.B
\end{equation}
where $x$ is not free in $B$ in the second scheme.
We show that inductive properties of $\lambda$-tree syntax 
specifications can be stated directly and in a purely logical
fashion, and proved in $\LGN.$

\paragraph{Relation to nominal logic}
In formulating the proof system for $\LGN$, it turns out that
we can simplify the presentation a lot if we adopt the idea of
{\em equivariant predicates} from nominal logic~\cite{pitts03ic}.
That is, provability of a predicate is invariant under
permutations of {\em names}. This is technically done by
introducing a countably infinite set of name constants
into the logic, and change the identity rule of the logic
to allow equivalence under permutations of name constants:
$$
\infer[id]
{\Seq {\Gamma, B}{B'} }
{\pi.B = \pi'.B'}
$$
where $\pi$ and $\pi'$ are permutations on names.
$\LGN$ is in fact very close to nominal logic, when we consider
only the behaviours of logical connectives. In particular,
the quantifier $\nabla$ in $\LGN$ shares the same properties,
in relation to other connectives of the logic, 
with the $\NEW$ quantifier in nominal logic. However, there are
two important differences in our approach.
First, we do not attempt to redefine $\alpha$-conversion and
substitutions in $\LGN$ in terms of permutations (or {\em swapping})
and the notion of {\em freshness} as in nominal logic.
Name swapping and freshness constraints are not part of the syntax
of $\LGN.$ These notions are present only in the meta theory of
the logic. 
In $\LGN$, for example, variables are always considered
to have empty support, that is, $\pi.x = x$ for every permutation $\pi$.
This is because we restrict substitutions to the ``closed'' ones,
in the sense that no name constants can appear in the substitutions.
A restricted form of open substitutions can be recovered indirectly 
at the meta theory of $\LGN$. 
The fact that variables have empty support 
allows one to work with permutation free formulas and terms. 
So in $\LGN$, we can prove that $p~x~a \oimp p~x~b$, where $a$ and
$b$ are names, without using explicit axioms of permutations
and freshness. In nominal logic, one would prove this by
using the swapping axiom $p~x ~ a \oimp p~((a~b).x)~((a~b).b)$, 
where $(a~b)$ denotes a swapping of $a$ and $b$,
and then show that $(a~b).x = x$. The latter might not be valid
if $x$ is substituted by $a$, for example. 
The validity of this formula in nominal logic 
would therefore depend on the assumption on the support of $x$.

The second difference between $\LGN$ and nominal logic is that 
$\LGN$ allows closed terms (again, in the sense that no name 
constants appear in them) of type
name, while in nominal logic, allowing such terms would
lead to an inconsistent theory in nominal logic~\cite{pitts03ic}.
As an example, the type $tm$ in the encoding of $\lambda$-terms
mentioned previously can be treated as a nominal type in $\LGN$.
This has an important consequence that we do not need to 
redefine the notion of substitutions for the encoded $\lambda$-terms.
For example, we can define the (lazy) evaluation relation on 
untyped $\lambda$-terms as the theory:
$$
\begin{array}{c}
eval ~ (abs ~ M) ~ (abs ~ M) \equiv \top \\
eval ~ (app~M~N) ~ V \equiv eval ~ M ~ (abs~P) \land eval ~ (P~N) ~ V
\end{array}
$$
without having to explicitly define substitutions on terms of type $tm$ 
inside $\LGN.$ Substitutions in the object language in this case 
is modelled by $\beta$-reduction in the meta-language of $\LGN.$

\paragraph{Outline of the paper} 
Section~\ref{sec:lgn} introduces the logic $\LG$,
which is an extension of first order intuitionistic logic with a notion of name permutation 
and the $\nabla$-quantifier. $\LG$ serves as the core logic for a more expressive 
logic, $\LGN$, which is obtained by adding rules for fixed points, equality and induction to $\LG.$
Section~\ref{sec:drv} examines several properties of derivations, in particular,
those that concern preservation of provability under several operations on
sequents, e.g., substitutions. 
Section~\ref{sec:reduct} defines the cut reduction, used in the cut-elimination proof.
The cut elimination proof itself is an adaptation of the cut-elimination proof
of $\FOLDN$ by McDowell and Miller~\cite{mcdowell00tcs}, which makes use of the reducibility
technique. Section~\ref{sec:norm} defines the normalizability and the reducibility
relations which are crucial to the cut elimination proof in Section~\ref{sec:cut-elim}.
Finally, in Section~\ref{sec:corr}, we show that the proof system $\LG$ is actually
equivalent to $\FOLDNb$ (without fixed points and equality) with non-logical rules
corresponding to the axioms given in (\ref{eq1}) above.

This paper contains the technical proofs for the 
results stated in \cite{tiu07entcs}; readers are encouraged to consult \cite{tiu07entcs} for
motivations and examples for $LG^\omega.$

\section{A logic for generic judgments}
\label{sec:lgn}

\newcommand{\swap}[2]{(#1 ~ #2)}

We first define the core fragment of the logic $\LGN$
which does not have fixed point rules or induction.
The starting point is the logic $\FOLNb$ introduced in ~\cite{miller05tocl}.
$\FOLNb$ is an extension of a subset of Church's 
Simple Theory of Types in which 
formulas are given the type $o$. The core fragment of $\LGN$,
which we refer to as $\LG$, shares the same set of connectives
as $\FOLNb$, namely, $\bot$, $\top$, $\land$, $\lor$,
$\oimp$, $\forall_\tau$, $\exists_\tau$ and $\nabla_\tau.$
The type $\tau$ in the quantifiers is restricted to that
which does not contain the type $o.$ Hence the logic is
essentially first-order. We abbreviate $(B \oimp C)
\land (C \oimp B)$ as $B \equiv C.$

The sequents of $\FOLDNb$ are expressions of the form
$$
\NSeq{\Sigma}{\sigma_1\triangleright B_1, \ldots, \sigma_n \triangleright B_n}
{\sigma_0 \triangleright B_0}
$$
where $\Sigma$ is a signature, i.e., a set of eigenvariables 
scoped over the sequent
and $\sigma_i$ is a local signature, i.e., 
list of variables locally scoped over $B_i$. The introduction rules
for $\nabla$, reading the rules bottom-up, introduce new local
variables to the local signatures, just as the right introduction rule
of $\forall$ introduces new eigenvariables to the signature.
The expression $\sigma_i \triangleright B_i$ is called a local judgment,
and is identified up to renaming of variables in $\sigma_i$. 
This enforces a limited notion of equivariance: 
for example $\Seq{a\triangleright p a}
{b \triangleright p b}$ is provable, since both local judgments
are equivalent up to renaming of local signatures. 
However, the judgments 
$(a,c)\triangleright p\, a$ and $b \triangleright p\, b$ are considered
distinct judgments, and so are $(a,b) \triangleright q\,a\,b$
and $(b,a)\triangleright q\,a\,b$. These restrictions are relaxed in $\LG.$

The sequent presentation of $\LG$ can be simplified,
that is, without using the local signatures, if we employ
the equivariance principle. 
For this purpose, we introduce a distinguished set of base types, 
called {\em nominal types}, which is denoted with ${\cal N}$. 
Nominal types are ranged over by $\iota$.
We restrict the $\nabla$ quantifier to nominal types. 
For each nominal type $\iota \in {\cal N}$,
we assume an infinite number of constants of that type.
These constants are called {\em nominal constants}.
We denote the family of nominal constants
by ${\cal C}_{\cal N}.$ 
The role of the nominal constants is to enforce the notion
of equivariance: provability of formulas
is invariant under permutations of nominal constans. 
Depending on the application, we might also assume a set of 
non-nominal constants, which is denoted by ${\cal K}.$

We assume the usual notion of capture-avoiding substitutions.
Substitutions are ranged over by $\theta$ and $\rho$. 
Application of substitutions is written in a postfix notation,
e.g., $t\theta$ is an application of $\theta$ to the term $t$.
Given two substitutions $\theta$ and $\theta'$, 
we denote their composition by $\theta \circ \theta'$
which is defined as $t(\theta \circ \theta') = (t\theta)\theta'.$
A {\em signature} is a set of variables. 
A substitution $\theta$ respects a given signature $\Sigma$
if there exists a set of typed variables $\Sigma'$
such that for every $x : \tau$ in the domain of $\theta$,
it holds that
$
\TSeq{{\cal K} \cup \Sigma' }
{\theta(x)}{\tau}.
$
We denote by $\Sigma\theta$ the minimal set of variables 
satisfying the above condition. 
We assume that variables, free or bound, are of a different
syntactic category from constants.

\begin{definition}
A permutation on ${\cal C}_{\cal N}$ is a bijection
from ${\cal C}_{\cal N}$ to ${\cal C}_{\cal N}$.
The permutations on ${\cal C}_{\cal N}$ are ranged over by $\pi$.
Application of a permutation $\pi$ to a nominal constant $a$
is denoted with $\pi(a)$. We shall be concerned only with
permutations which respect types, i.e., 
for every $a:\iota$,  $\pi(a) : \iota.$
Further, we shall also restrict to permutations which
are finite, that is, the set
$\{ a \mid \pi(a) \not = a \}$ is finite. 
Application of a permutation to an arbitrary term
(or formula), written $\pi.t$, is defined as follows:
$$
\begin{array}{ccc}
\pi.a  =  \pi(a), \hbox{ if $a \in {\cal C}_{\cal N}.$ } & 
\pi.c  =  c, \quad \hbox{if } c \not \in {\cal C}_{\cal N}. &
\pi.x  =  x  \\
\pi.(M~N)  =  (\pi.M) ~ (\pi.N) & \quad
\pi.(\lambda x.M)  =  \lambda x. (\pi.M) & 
\end{array}
$$
A permutation involving only two nominal constants
is called {\em swapping}. We use $(a~b)$, where
$a$ and $b$ are constants of the same type, to denote
the swapping $\{ a \mapsto b, b \mapsto a\}.$
\end{definition}

The {\em support} of a term (or formula) $t$, written $supp(t)$, is the set of
nominal constants appearing in it. 
It is clear from the above definition that if $supp(t)$ is
empty, then $\pi.t = t$ for all $\pi$.
The definition of $\Sigma$-substitution implies that
for every $\theta$ and for every $x \in dom(\theta)$, 
$\theta(x)$ has empty support.
Therefore $\Sigma$-substitutions and permutations commute, 
that is, $(\pi.t)\theta = \pi.(t\theta).$

A sequent in $\LGN$ is an expression of the form
$
\NSeq \Sigma \Gamma C
$
where $\Sigma$ is a signature. The free variables
of $\Gamma$ and $C$ are among the variables in $\Sigma$.
The inference rules for the core fragment of $\LGN$, i.e.,
the logic $\LG$, is given in Figure~\ref{fig:LG}.
In the rules, the typing judgment
$
\TSeq{\Sigma, {\cal K}, {\cal C}_{\cal N}}
{t}{\tau}
$
denotes the typability of $t : \tau$, given the typing
context $\Sigma \cup {\cal K} \cup {\cal C}_{\cal N}$
in Church's simple type system.

In the $\nablaL$ and $\nablaR$ rules, $a$ denotes
a nominal constant.
In the $\existsL$ and $\forallR$ rules, we use 
{\em raising} \cite{miller92jsc} 
to encode the dependency of the quantified variable
on the support of $B$, since we do not allow
$\Sigma$-substitutions to mention any nominal constants. 
In the rules, the variable $h$ has its type raised
in the following way: suppose $\vec c$
is the list $c_1:\iota_1,\dots,c_n:\iota_n$ and
the quantified variable $x$ is of type $\tau$. Then
the variable $h$ is of type:
$\iota_1 \to \iota_2 \to \dots \to \iota_n \to \tau.$ 
This raising technique is similar to that of $\FOLDNb,$
and is used to encode explicitly the minimal 
support of the quantified variable. 
Its use prevents one from mixing the scopes of $\forall$ (dually, $\exists$)
and $\nabla$. That is, it prevents the formula 
$\forall x \nabla y. p\,x\,y \equiv \nabla y\forall x. p\,x\,y$,
and its dual, to be proved.

Looking at the introduction rules for $\forall$ and $\exists$,
one might notice the asymmetry between the
left and the right introduction rules. 
The left rule for $\forall$ allows instantiations with terms
containing any nominal constants 
while the raised variable in the right introduction rule of $\forall$
takes into account only those which are in the support of the quantified
formula.
However, we will see that we can extend the dependency of the 
raised variable to an arbitrary number of fresh nominal constants
not in the support without affecting the provability of the sequent
(see Lemma~\ref{lm:supp1} and Lemma~\ref{lm:supp2}).

\begin{figure}
{\small
$$
\infer[id_\pi]
{\NSeq \Sigma {\Gamma,B} {B'}}
{\pi.B = \pi'.B'}
\qquad
\infer[\mc]
{\NSeq{\Sigma}{\Delta_1,\ldots, \Delta_n,\Gamma}{C}}
{\NSeq{\Sigma}{\Delta_1}{B_1} & \cdots & \NSeq{\Sigma}{\Delta_n}{B_n}
& \NSeq{\Sigma}{B_1,\dots,B_n,\Gamma}{C}}
\qquad
\infer[\cL]
{\NSeq \Sigma {\Gamma, B} C}
{\NSeq \Sigma{\Gamma, B, B} C}
$$
$$
\infer[\botL]
{\NSeq \Sigma {\Gamma, \bot} {C}}
{}
\qquad
\infer[\topR]
{\NSeq \Sigma \Gamma \top}
{}
$$
$$
\infer[\landL, i \in \{1,2\}]
{\NSeq{\Sigma}{\Gamma, B_1 \land B_2}{C}}
{\NSeq{\Sigma}{\Gamma, B_i}{C}}
\qquad
\infer[\landR]
{\NSeq{\Sigma}{\Gamma}{B \land C}} 
{\NSeq \Sigma  \Gamma B & \NSeq \Sigma \Gamma C}
$$
$$
\infer[\lorL]
{\NSeq \Sigma  {\Gamma, B \lor D} C}
{\NSeq \Sigma  {\Gamma, B} C 
& \NSeq \Sigma  {\Gamma, D} C}
\qquad
\infer[\lorR, i \in \{1,2\}]
{\NSeq \Sigma  \Gamma {B_1 \lor B_2}}
{\NSeq \Sigma  \Gamma {B_i}}
$$
$$
\infer[\oimpL]
{\NSeq \Sigma  {\Gamma, B \oimp D} C }
{\NSeq \Sigma  \Gamma B & \NSeq \Sigma  {\Gamma, D} C}
\qquad
\infer[\oimpR]
{\NSeq \Sigma  \Gamma {B \oimp C}}
{\NSeq \Sigma {\Gamma, B} C}
$$
$$
\infer[\forallL]
{\NSeq \Sigma  {\Gamma, \forall_\tau x. B} C}
{\TSeq {\Sigma, {\cal K}, {\cal C}_{\cal N}} t \tau & \NSeq \Sigma {\Gamma, B[t/x]} C}
\qquad
\infer[\forallR, h \not \in \Sigma, supp(B) = \{\vec c\}]
{\NSeq \Sigma  \Gamma {\forall x.B}}
{\NSeq {\Sigma,h} \Gamma {B[h\, \vec c/x]} }
$$
$$
\infer[\nablaL, a \not \in supp(B)]
{\NSeq \Sigma {\Gamma, \nabla x.B} C}
{\NSeq \Sigma {\Gamma, B[a/x]} C}
\qquad
\infer[\nablaR, a \not \in supp(B)]
{\NSeq \Sigma \Gamma {\nabla x.B}}
{\NSeq \Sigma \Gamma {B[a/x]}}
$$
$$
\infer[\existsL, h \not \in \Sigma, supp(B) = \{\vec c\}]
{\NSeq \Sigma  {\Gamma, \exists x.B} C}
{\NSeq {\Sigma, h}  {\Gamma, B[h\,\vec c/x]} C}
\qquad
\infer[\existsR]
{\NSeq \Sigma  \Gamma {\exists_\tau x.B}}
{\TSeq {\Sigma,{\cal K},{\cal C}_{\cal N}} t \tau & \NSeq \Sigma \Gamma {B[t/x]}}
$$
}
\caption{The inference rules of $\LG$}
\label{fig:LG}
\end{figure}

We now extend the logic $\LG$ with a proof theoretic notion
of equality and fixed points, following on works by
Hallnas and Schroeder-Heister~\cite{hallnas91jlc,schroeder-heister92nlip}, 
Girard~\cite{girard92mail} and McDowell and Miller~\cite{mcdowell00tcs}.
The equality rules are as follows:
$$
\infer[\eqL]
{\NSeq \Sigma {\Gamma, s = t} C}
{
\{\NSeq {\Sigma\theta}  {\Gamma\theta} {C\theta}
~ \mid ~ (\lambda \vec c.t)\theta =_{\beta\eta} (\lambda \vec c.s)\theta  \}
}
\qquad
\infer[\eqR]
{\NSeq \Sigma \Gamma {t = t}}
{}
$$
where $supp(s = t) = \{\vec c\}$ in the $\eqL$ rule.
In the $\eqL$ rule, the substitution $\theta$ is a {\em unifier} of
$\lambda \vec c.s$ and $\lambda \vec c.t$.
We specify the premise of the rule as a set to mean that every element
of the set is a premise. Since the terms $s$ and $t$ can be
arbitrary higher-order terms, in general the set of their unifiers 
can be infinite. However, in some restricted cases, 
e.g., when $\lambda \vec c.s$ and $\lambda \vec c.t$ are 
{\em higher-order pattern} terms~\cite{miller91jlc,nipkow93lics},
if both terms are unifiable, then there exists a most general unifier.
The applications we are considering are those which satisfy the 
higher-order pattern restrictions.

\begin{definition}
To each atomic formula, we associate a fixed point equation,
or a {\em definition clause}, following the terminology of
$\FOLDNb$. A definition clause is written
$
\forall \vec x. p\,\vec x \defeq B
$
where the free variables of $B$ are among $\vec x.$
The predicate $p\,\vec x$ is called the {\em head}
of the definition clause, and $B$ is called the {\em body}.
A {\em definition} is a set of definition clauses. 
We often omit the outer quantifiers when referring to
a definition clause.
\end{definition}

The introduction rules for defined atoms are as follows:
$$
\infer[\defL, p\,\vec x \defeq B]
{\NSeq \Sigma {\Gamma, p\, \vec t} C}
{\NSeq \Sigma {\Gamma, B[\vec t/\vec x]} C}
\qquad
\infer[\defR, p\,\vec x \defeq B]
{\NSeq \Sigma  \Gamma {p\, \vec t}}
{\NSeq \Sigma  \Gamma {B[\vec t/\vec x]}}
$$

In order to prove the cut-elimination theorem and
the consistency of $\LGN$, we allow only definition clauses 
which satisfy an {\em equivariance preserving} condition and 
a certain positivity condition, so as to guarantee the
existence of fixed points. 

\newcommand{\level}[1]{lvl(#1)}
\begin{definition}
\label{def:level}
We associate with each predicate symbol $p$ a natural number,
the {\em level} of $p$. 
Given a formula $B$, its {\em level} $\level{B}$ is defined as follows:
\begin{enumerate}
\item $\level{p \, \bar{t}} = \level{p}$
\item $\level{\bot} = \level{\top} = 0$
\item $\level{B \land C} = \level{B \lor C} = \max(\level{B},\level{C})$
\item $\level{B \oimp C} = \max(\level{B}+1,\level{C})$
\item $\level{\forall x.B} = \level{\nabla x.B} = \level{\exists x.B}
       = \level{B}$. 
\end{enumerate}
A definition clause $p\,\vec x \defeq B$ is stratified if
$\level{B} \leq \level{p}$ and
$B$ has no free occurrences of nominal constants.
We consider only definition clauses which are stratified.
\end{definition}
An example that violates the first restriction in 
Definition~\ref{def:level} is the definition $p \defeq p \oimp \bot.$
In~\cite{schroeder-heister92nlip}, Schroeder-Heister shows that admitting this
definition in a logic with contraction leads to
inconsistency. To see why we need the second restriction
on name constants, consider the definition $q\, x \defeq (x = a),$
where $a$ is a nominal constant. Let $b$ be a
nominal constant different from $a$. 
Using this definition, we would be able to derive $\bot$:
$$
\infer[cut]
{\Seq{}{\bot}}
{
   \infer[\defR]
   {\Seq{}{q\,a}}
   {
      \infer[\eqR]
      {\Seq{}{a = a}}
      {}
   }
&
   \infer[cut]
   {\Seq{q\,a}{\bot}}
   {
     \infer[id_\pi]
     {\Seq{q\,a}{q\,b}}
     {}
   &
     \infer[\defL]
     {\Seq{q\,b}{\bot}}
     {
       \infer[\eqL]
       {\Seq{b = a}{\bot}}
       {}
     }
   }
}
$$

In examples and applications, we
often express definition clauses with patterns in the heads. 
Let us consider, for example, a definition clause for lists. 
We first introduce a type $lst$ to denote lists of elements of type $\alpha$, 
and the constants
$$
nil : lst \qquad :: ~ : \alpha \to lst \to lst
$$
which denote the empty list and a constructor to build a list
from an element of type $\alpha$ and another list. The latter
will be written in the infix notation.   
The definition clause for {\em lists} is as follows.
$$
list~L \defeq L = nil \lor \exists_\alpha A\exists_{lst} L'. 
     L = (A::L') \land list~L'.
$$
Using patterns, the above definition of lists can
be rewritten as
$$
list~nil \defeq \top.\qquad \quad
list~(A :: L) \defeq list~L.
$$

We shall often work directly with this patterned notation for definition clauses.
For this purpose, we introduce the notion of {\em patterned definitions}.
A {\em patterned definition clause} is written
$\forall \vec x. H \defeq B$ where the free variables of $H$  and $B$ are among
$\vec x.$ The stratification of definitions in Definition~\ref{def:level}
applies to patterned definitions as well. 
Since the patterned definition clauses are not allowed to have
free occurrences of nominal constants, in matching the heads of the
clauses with an atomic formula in a sequent, we need to raise the variables of
the clauses to account for nominal constants that are in the support
of the introduced formula.
Given a patterned definition clause 
$
\forall x_1 \dots \forall x_n. H \defeq B
$
its raised clause with respect to the list of
constants $c_1:\iota_1 \dots c_n:\iota_n$
is
$$
\forall h_1 \dots \forall h_n. H[h_1~\vec c/x_1, \ldots, h_n~\vec c/x_n] 
\defeq B[h_1~\vec c/x_1, \ldots, h_n~\vec c/x_n].
$$
The introduction rules for patterned definitions are
$$
\infer[\defL]
{\NSeq{\Sigma}{A, \Gamma}{C}}
{\{\NSeq{\Sigma\theta}{B\theta, \Gamma\theta}{C\theta}\}_\theta}
\qquad
\infer[\defR]
{\NSeq{\Sigma}{\Gamma}{A}}
{\NSeq{\Sigma}{\Gamma}{B\theta}}
$$
In the $\defL$ rule, $B$ is the body of the raised patterned clause
$\forall x_1 \dots \forall x_n. H \defeq B$ and
$(\lambda \vec c.H)\theta = (\lambda \vec c.A)\theta$ where $\{\vec c\}$ is
the support of $A.$ 
In the $\defR$ rule, we match $A$ with the head of the clause,
i.e., $\lambda \vec c.A = (\lambda \vec c.H)\theta.$
These patterned rules can be derived using the non-patterned
definition rules and the equality rules, as shown in~\cite{tiu04phd},

\paragraph{Natural number induction.}
We introduce a type $nt$ to denote natural numbers,
with the usual constants $z : nt$ (zero) and $s : nt \to nt$ (the
successor function), and a special predicate $nat : nt \to nt \to o.$
The rules for natural number induction are the same as
those in $\FOLDN$~\cite{mcdowell00tcs}, which are the introduction rules
for the predicate $nat$.
$$
\infer[nat{\cal L}]
{\NSeq{\Sigma}{\Gamma, nat\,I}{C}}
{\Seq{}{D\,z} & \NSeq{j}{D\,j}{D\,(s\,j)} 
& \NSeq{\Sigma}{\Gamma, D\,I}{C}}
$$

$$
\infer[nat{\cal R}]
{\NSeq{\Sigma}{\Gamma}{nat~z}}
{}
\qquad
\infer[nat{\cal R}]
{\NSeq{\Sigma}  \Gamma {nat\,(s\,I)}}
{\NSeq{\Sigma}  \Gamma {nat\,I}}
$$

The logic $\LG$ extended with the equality, definitions and
induction rules is referred to as $\LGN.$

\section{Properties of derivations}
\label{sec:drv}

In this section we examine several properties of the $\nabla$-quantifier and 
derivations in $\LGN$ that are useful in the cut elimination proof. 
These properties concern the transformation of derivations, in particular,
they state that provability is preserved under $\Sigma$-substitutions, permutations
and a restricted form of name substitutions.

We first look at the properties of the $\nabla$ quantifier
in relation to other connectives. 
The proof of the following proposition is straightforward by
inspection on the rules of $\LG.$

\begin{proposition}
The following formulas are provable in $\LG$:
\begin{enumerate}
\item $\nabla x.(B x \land C x) \equiv \nabla x.B x \land \nabla x. C x.$
\item $\nabla x.(B x \oimp C x) \equiv \nabla x.B x \oimp \nabla x. C x.$
\item $\nabla x.(B x \lor C x) \equiv \nabla x.B x \lor \nabla x.C x.$
\item $\nabla x. B \equiv B$, provided that $x$ is not free in $B$.
\item $\nabla x\nabla y.B x y \equiv \nabla y\nabla x.Bxy.$
\item $\forall x.B x \oimp \nabla x.B x.$
\item $\nabla x.B x \oimp \exists x.B x.$ 
\end{enumerate}
\end{proposition}
The formulas (1) -- (3) are provable in $\FOLNb$. The proposition
is true also in nominal logic with $\nabla$ replaced by $\NEW.$

\begin{definition}
\label{def:ht}
Given a derivation $\Pi$ with premise derivations 
$\{\Pi_i\}_{i \in {\cal I}}$ where ${\cal I}$ is some
index set, the measure $ht(\Pi)$, the {\em height} of $\Pi$, 
is defined as the least upper bound of
$
\{ht(\Pi_i) + 1\}_{i \in {\cal I}}.
$
\end{definition}

We now define some transformations of derivations:
weakening of hypotheses, substitutions on derivations,
permutations and restricted name substitutions.
In the following definitions we omit the signatures
in the sequents if it is clear from context which
signatures we refer to.
We denote with $id$ the identity function on ${\cal C}_{\cal N}$.

\begin{definition}
\label{def:weak}
{\em Weakening of hypotheses.}
Let $\Pi$ be a derivation of $\NSeq{\Sigma}{\Gamma}{C}.$
Let $\Delta$ be a multiset of formulas whose free variables
are among $\Sigma$. We define the derivation $w(\Delta,\Pi)$
of $\NSeq{\Sigma}{\Gamma, \Delta}{C}$ as follows:
\begin{enumerate}
\item If $\Pi$ ends with $\eqL$
$$
\infer[\eqL]
{\NSeq{\Sigma}{s = t, \Gamma'}{C}}
{\left\{
\raisebox{-1.5ex}{\deduce{\NSeq{\Sigma\theta}{\Gamma'\theta}{C\theta}}{\Pi_\theta}}
\right\}_\theta
}
$$
then $w(\Delta, \Pi)$ is 
$$
\infer[\eqL]
{\NSeq{\Sigma}{s = t, \Gamma', \Delta}{C}}
{
\left\{
\raisebox{-1.5ex}{\deduce{\NSeq{\Sigma\theta}{\Gamma'\theta, \Delta\theta}{C\theta}}
{w(\Delta\theta,\Pi_\theta)}}
\right\}_\theta
}
$$
\item If $\Pi$ ends with $\natL$
$$
\infer[\natL]
{\Seq{nat~I, \Gamma'}{C}}
{\deduce{\Seq{}{D~z }}{\Pi_1}
& \deduce{\Seq{D~i}{D~(s~i)}}{\Pi_2}
& \deduce{\Seq{D~I, \Gamma'}{C}}{\Pi_3}}
$$
then $w(\Delta, \Pi)$ is
$$
\infer[\natL]
{\Seq{nat~I, \Gamma', \Delta}{C}}
{\deduce{\Seq{}{D~z }}{\Pi_1}
& \deduce{\Seq{D~i}{D~(s~i)}}{\Pi_2}
& \deduce{\Seq{D~I, \Gamma', \Delta}{C}}{w(\Delta,\Pi_3)}}
$$
\item If $\Pi$ ends with the $mc$ rule
$$
\infer[mc]
{\Seq{\Delta_1,\ldots,\Delta_n, \Gamma'}{C}}
{
\deduce{\Seq{\Delta_1}{B_1}}{\Pi_1}
& \ldots &
\deduce{\Seq{\Delta_n}{B_n}}{\Pi_n}
& 
\deduce{\Seq{B_1,\dots,B_n,\Gamma'}{C}}{\Pi'}
}
$$
then $w(\Delta,\Pi)$ is
$$
\infer[mc]
{\Seq{\Delta_1,\ldots,\Delta_n, \Gamma', \Delta}{C}}
{
\deduce{\Seq{\Delta_1}{B_1}}{\Pi_1}
& \ldots &
\deduce{\Seq{\Delta_n}{B_n}}{\Pi_n}
& 
\deduce{\Seq{B_1,\dots,B_n,\Gamma', \Delta}{C}}{w(\Delta,\Pi')}
}
$$
\item If $\Pi$ ends with any other rule and has premise derivations
$\Pi_1,\dots,\Pi_n$ then $w(\Delta,\Pi)$ ends with the same
rule with premise derivations $w(\Delta,\Pi_n),$ $\dots,$ $w(\Delta,\Pi_n).$
\end{enumerate}
\end{definition}

\begin{definition}
\label{def:subst}
{\em Substitutions on derivations.}
If $\Pi$ is a derivation of $\NSeq{\Sigma}{\Gamma}{C}$ and $\theta$
is a $\Sigma$-substitution, then we define the derivation $\Pi\theta$ 
of $\NSeq{\Sigma\theta}{\Gamma\theta}{C\theta}$ as follows:
\begin{enumerate}
\item Suppose $\Pi$ ends with $\eqL$:
$$
\infer[\eqL]
{\NSeq{\Sigma}{s = t, \Gamma'}{C}}
{\left\{
\raisebox{-1.5ex}{\deduce{\NSeq{\Sigma\rho}{\Gamma'\rho}{C\rho}}{\Pi_\rho}}
\right\}_\rho
}
$$
where each $\rho$ is a unifier of $\lambda \vec c.s$
and $\lambda \vec c.t$. Observe that if $\rho'$ is a unifier
of $(\lambda \vec c.s)\theta$ and $(\lambda \vec c.t)\theta$, 
then $\theta\circ \rho'$ is a unifier of  $\lambda \vec c.s$
and $\lambda \vec c.t$. Thus $\Pi\theta$ is the derivation:
$$
\infer[\eqL]
{\NSeq{\Sigma}{s\theta = t\theta, \Delta\theta}{C\theta}}
{
\left\{ 
\raisebox{-1.5ex}{\deduce{\NSeq{\Sigma\theta\rho'}{\Delta\theta\rho}{C\theta\rho}}
{\Pi_{\theta\circ \rho'}}}
\right\}_{\rho'}
}
$$
\item Suppose $\Pi$ ends with $\forallR$:
$$
\infer[\forallR]
{\NSeq{\Sigma}{\Gamma}{\forall x.B}}
{\deduce{\NSeq{\Sigma}{\Gamma}{B[h\,\vec c/x]}}{\Pi_1}}
\enspace ,
$$
where $\{\vec c\} = supp(\forall x.B).$
Let $\{\vec d\}$ be the support of $(\forall x.B) \theta$,
which might be smaller than $\{ \vec c\}.$
Let $\rho$ be the substitution $[\lambda \vec c.h' \vec d/h]$
where $h'$ is a new variable not already in $\Sigma$ and 
not among the free variables in $\theta.$ 
We can assume without loss of generality that 
$x$ is not free in $\theta$, hence 
$((B[h\,\vec c/x])\rho) \theta = (B[h'\,\vec d/x])\theta = (B\theta)[h'\,\vec d/x].$
Then $\Pi\theta$ is
$$
\infer[\forallR]
{\NSeq{\Sigma\theta}{\Gamma\theta}{(\forall x.B)\theta}}
{\deduce{\NSeq{\Sigma\theta, h'}{\Gamma\theta}{(B\theta)[h'\,\vec d/x]}}{\Pi_1(\rho \circ \theta)}}
$$
\item Suppose $\Pi$ ends with $\existsL$: this case is dual to the previous one.
\item If $\Pi$ ends with any other rule and has premise derivations
$\Pi_1,\dots,\Pi_n$, then $\Pi\theta$ ends with the same rule and has
premise derivations $\Pi_1\theta$,$\dots,$ $\Pi_n\theta.$
\end{enumerate}
\end{definition}

\begin{definition}
\label{def:perm}
Let $\Pi$ be a proof of $\NSeq \Sigma {B_1, \ldots, B_n} {B_0}$
and let $\vec \pi = \pi_0,\dots,\pi_n$ be a list of permutations. We define
a derivation $\langle\vec \pi\rangle.\Pi$ of 
$\NSeq{\Sigma}{\pi_1.B_1, \ldots, \pi_n.B_n}{\pi_0.B_0}$
as follows:
\begin{enumerate}
\item Suppose that $\Pi$ ends with $id_\pi$
$$
\infer[id_\pi]
{\NSeq \Sigma {B_1, \ldots, B_n}{B_0}}
{\pi.B_j = \pi'.B_0}
\enspace .
$$
Obverse that $\pi.\pi_j^{-1}.\pi_j.B = \pi'.\pi_0^{-1}.\pi_0.B'.$
Hence $\langle\vec \pi\rangle.\Pi$ ends with the same rule.

\item Suppose $\Pi$ ends with $mc$:
$$
\infer[mc]
{\Seq{B_1,\ldots,B_n}{B_0}}
{
\deduce{\Seq{\Delta_1}{D_1}}{\Pi_1}
&
\ldots
&
\deduce{\Seq{\Delta_m}{D_m}}{\Pi_m}
&
\deduce{\Seq{D_1,\ldots,D_m,\Delta_{m+1}}{B_0}}{\Pi'}
}
$$
where $\Delta_1, \ldots, \Delta_{m+1}$ are partitions of
$B_1,\dots,B_n.$ Suppose that for each $i \in \{1,\dots,m+1\}$, 
$\Delta_i = B_{i1},\dots,B_{ik_i}$ for some index $k_i.$
Let $\vec \pi(i)$, for $i \in \{1,\dots,m\}$, 
be the permutations $id,\pi_{i1},\dots,\pi_{ik_i}.$
Let $\vec \pi(m+1)$ be the permutations
$$\pi_0, \underbrace{id,\dots,id}_m,\pi_{(m+1)1},\ldots \pi_{(m+1)k_{m+1}}$$

We denote with $\Delta_i'$ the list
$$
\pi_{i1}.B_{ij}, \ldots, \pi_{ik_i}.B_{ik_i}.
$$
Then $\langle \vec \pi \rangle.\Pi$ is the derivation
$$
\infer[mc]
{\Seq{\pi_1.B_1,\ldots,\pi_n.B_n}{\pi_0.B_0}}
{
\deduce{\Seq{\Delta_1'}{D_1}}{\langle \vec \pi(1)\rangle.\Pi_1}
&
\ldots
&
\deduce{\Seq{\Delta_m'}{D_m}}{\langle \vec \pi(m)\rangle.\Pi_m}
&
\deduce{\Seq{D_1,\ldots,D_m,\Delta_{m+1}'}{\pi_0.B_0}}{\langle \vec \pi(m+1)\rangle.\Pi'}
}
$$

\item Suppose $\Pi$ ends with $\nablaR$:
$$
\infer[\nablaR]
{\NSeq{\Sigma}{B_1,\ldots,B_n}{\nabla_\iota x.B}}
{\deduce{\NSeq{\Sigma}{B_1,\ldots,B_n}{B[a/x]}}{\Pi_1}}
$$
where $a : \iota \not \in supp(B).$ 
Let $d : \iota$ be a nominal constant such that 
$d \not \in supp(B)$ and $\pi_0(d) = d$. Such a constant
exists since $supp(B)$ is finite and $\pi_0$ is a finite
permutation. Thus $\pi_0.(a~d).B_0[a/x] = \pi_0.B_0[d/x].$
Then $\langle\vec \pi\rangle.\Pi$ is the derivation:
$$
\infer[\nablaR]
{\NSeq{\Sigma}{\pi_1.B_1, \ldots, \pi_n.B_n}{\pi_0.(\nabla x.B)}}
{\deduce{\NSeq{\Sigma}{\pi_1.B_1, \ldots, \pi_n.B_n}{\pi_0.B[d/x]}}
{\langle\pi_0.(a~d), \dots,\pi_n\rangle.\Pi_1}}
$$
\item Suppose $\Pi$ ends with $\nablaL$: this case is analogous to previous one.

\item Suppose $\Pi$ ends with $\cL$:
$$
\infer[\cL]
{\Seq{B_1,\ldots,B_j,\ldots,B_n}{B_0}}
{
\deduce{
\Seq{B_1,\ldots,B_j,B_j\ldots,B_n}{B_0}
}
{
\Pi'
}
}
$$
then $\langle\vec \pi\rangle.\Pi$ is
$$
\infer[\cL]
{\Seq{\pi_1.B_1,\ldots,\pi_{j}.B_j,\ldots,\pi_n.B_n}{\pi_0.B_0}}
{
\deduce{
\Seq{\pi_1.B_1,\ldots,\pi_{j}.B_j,\pi_j.B_j\ldots,\pi_n.B_n}{\pi_0.B_0}
}
{
\langle \pi_1,\ldots,\pi_j,\pi_j,\ldots,\pi_n\rangle.\Pi'
}
}
$$
\item If $\Pi$ ends with any other rule and has premise derivations 
$\Pi_1, \dots, \Pi_m$, then $\langle \vec \pi \rangle.\Pi$ ends with
the same rule and has premise derivations
$\langle \vec \pi \rangle.\Pi_1,$ $\dots,$ $\langle \vec \pi \rangle.\Pi_m.$

\end{enumerate}
\end{definition}

\begin{definition}
\label{def:res}
Let $\Pi$ be a proof of $\NSeq{\Sigma,x:\iota}{B_1, \ldots, B_n}{B_0}$
and let $\vec a = a_0, \dots, a_n$ be a list of nominal constants
such that $a_i\not \in supp(B_i).$
We define a derivation $r(x,\langle \vec a\rangle, \Pi)$
of 
$\NSeq{\Sigma}{B_1[a_1/x], \ldots, B_n[a_n/x]}{B_0[a_0/x]},$
as follows:
\begin{enumerate}
\item Suppose $\Pi$ is
$$
\infer[id_\pi]
{\NSeq{\Sigma,x}{B_1,\ldots,B_n}{B_0}}
{\pi.B_j = \pi'.B_0}
\enspace .
$$
Let $d : \iota$ be a nominal constant which is not in
the support of $B_j$ and $B_0$, and $\pi(d) = d$
and $\pi'(d) = d$. 
Then $r(x,\vec a,\Pi)$ is 
$$
\infer[id_\pi]
{\NSeq{\Sigma}{B_1[a_1/x], \ldots, B_n[a_n/x]}{B_0[a_0/x]}}
{\pi.(a_j~d).B_1[a_1/x] = \pi'.(a_0~d).B_0[a_0/x]}
$$

\item Suppose $\Pi$ ends with $mc$:
$$
\infer[mc]
{\NSeq{\Sigma,x}{B_1,\ldots,B_n}{B_0}}
{
\deduce{\NSeq{\Sigma,x}{\Delta_1}{D_1}}{\Pi_1}
&
\ldots
&
\deduce{\NSeq{\Sigma,x}{\Delta_m}{D_m}}{\Pi_m}
&
\deduce{\NSeq{\Sigma,x}{D_1,\ldots,D_m,\Delta_{m+1}}{B_0}}{\Pi'}
}
$$
where $\Delta_1, \ldots, \Delta_{m+1}$ is a partition of
$B_1,\dots,B_n.$ Suppose that for each $i \in \{1,\dots,m+1\}$, 
$\Delta_i = B_{i1},\dots,B_{ik_i}$ for some index $k_i.$
Let $\vec d = d_1,\dots,d_m$ be a list of nominal constants
such that $d_i \not \in supp(D_i).$
Let $f(i)$, for $i \in \{1,\dots,m\}$ be the list
$d_i,a_{i1},\dots,a_{ik_i}$
and let $f(m+1)$ be the list
$$a_0,\vec d,a_{(m+1)1},\dots,a_{(m+1)k_{(m+1)}}.$$
Let $\Delta_i'$ be the list 
$$
B_{i1}[a_{i1}/x],\ldots,B_{ik_i}[a_{ik_i}/x]
$$
and let $\Gamma$ be the list
$$
D_{1}[d_1/x],\ldots,D_{m}[d_m/x], \Delta_{m+1}'.
$$
Then $r(x,\vec a, \Pi)$ is the derivation
$$
\infer[mc]
{\NSeq{\Sigma}{B_1[a_1/x],\ldots,B_n[a_n/x]}{B_0[a_0/x]}}
{
\deduce{\NSeq{\Sigma}{\Delta_1'}{D_1[d_1/x]}}{r(x,f(1),\Pi_1)}
&
\ldots
&
\deduce{\NSeq{\Sigma}{\Delta_m'}{D_m[a_m/x]}}{r(x,f(m),\Pi_m)}
&
\deduce{\NSeq{\Sigma}{\Gamma}{B_0[a_0/x]}}{r(x,f(m+1),\Pi')}
}
$$
\item Suppose $\Pi$ is
$$
\infer[\nablaR]
{\NSeq{\Sigma,x}{B_1, \ldots, B_n}{\nabla y.B}}
{\deduce{\NSeq{\Sigma,x}{B_1,\ldots,B_n}{B[c/y]}}{\Pi_1}}
\enspace .
$$
If $a_0 \not = c$ then $r(x,\vec a, \Pi)$ is
$$
\infer[\nablaR]
{\NSeq{\Sigma,x}{B_1, \ldots, B_n}{\nabla y.B}}
{\deduce{\NSeq{\Sigma,x}{B_1,\ldots,B_n}{B[c/y]}}{r(x,\vec a, \Pi_1)}}
\enspace .
$$
If $a_0 = c$, then we swap $c$ with a fresh constant.
Let $d : \iota$ be a nominal constant not in the support of $B[c/y]$.
We apply the swapping $(c~d)$ to the conclusion of the 
end sequent of $\Pi_1$ according to the construction
in Definition~\ref{def:perm} to get a proof $\Pi_2$
of $\NSeq{\Sigma,x}{B_1,\ldots,B_n}{B_0[d/y]}.$ 
The derivation $r(x,\vec a,\Pi)$ is constructed
as follows:
$$
\infer[\nablaR]
{\NSeq{\Sigma}{B_1[a_1/x],\ldots,B_n[a_n/x]}{\nabla y.B[a_0/x]}}
{\deduce{\NSeq{\Sigma}{B_1[a_1/x],\ldots,B_n[a_n/x]}{B[a_0/x, d/y]}}{r(x,\vec a,\Pi_2)}}
$$

\item If $\Pi$ ends with $\nablaL$ apply the same construction as in the
previous case.

\item Suppose $\Pi$ ends with $\forallR$
$$
\infer[\forallR]
{\NSeq{\Sigma,x}{B_1,\ldots,B_n}{\forall y.B}}
{\deduce{\NSeq{\Sigma,x,h}{B_1,\ldots,B_n}{B[h\,\vec c/y]}}{\Pi_1}}
\enspace .
$$
Let $\theta = [\lambda \vec c.h'\,\vec c x/h]$
where $h'$ is a variable not in $\Sigma.$
Apply the construction in Definition~\ref{def:subst}
to get the proof $\Pi\theta$ of 
$$
\NSeq{\Sigma,x,h'}{B_1,\ldots,B_n}{B[h'\,\vec a x/y]}
$$
Then $r(x,\vec a, \Pi)$ is 
$$
\infer[\forallR]
{\NSeq{\Sigma}{B_1[a_1/x],\ldots,B_n[a_n/x]}{\forall y.B[a_0/x]}}
{\deduce{\NSeq{\Sigma,h'}{B_1[a_1/x],\ldots,B_n[a_n/x]}
  {B[a_0/x, (h'\,\vec c a_0)/y]}}{r(x,\vec a, \Pi\theta)}
}
\enspace .
$$
\item If $\Pi$ ends with $\existsL$, apply the same construction as in
the previous case.

\item Suppose $\Pi$ ends with $\existsR$:
$$
\infer[\existsR]
{\NSeq{\Sigma,x}{B_1,\ldots,B_n}{\exists y.B}}
{\deduce{\NSeq{\Sigma,x}{B_1,\ldots,B_n}{B[t/y]}}{\Pi_1}}
\enspace .
$$
If $a_0\not \in supp(B[t/y])$ then $r(x,\vec a, \Pi)$ is
$$
\infer[\existsR]
{\NSeq{\Sigma}{B_1[a_1/x], \ldots, B_n[a_n/x]}{\exists y.B[a_0/x]}}
{\deduce{\NSeq{\Sigma}{B_1[a_1/x], \ldots, B_n[a_n/x]}{B[a_0/x,t/y]}}{r(x,\vec a,\Pi_1)}}
\enspace .
$$
If $a_0 \in supp(B[t/y]$, we exchange it with a fresh constant. 
Let $d$ be a nominal constant distinct from $a_0$ and not in the
support of $B[t/y].$ Then $((a_0~d).B[t/y])[a_0/x] = B[(a_0~d).t/y,a_0/x].$
We first apply the construction in Definition~\ref{def:perm}
to $\Pi_1$ to get a derivation $\Pi_2$ of 
$\NSeq{\Sigma,x}{B_1,\ldots,B_n}{B[(a_0~d).t/y, a_0/x]}.$
The derivation $r(x,\vec a, \Pi)$ is thus
$$
\infer[\existsR]
{\NSeq{\Sigma}{B_1[a_1/x],\ldots,B[a_n/x]}{\exists y.B[a_0/x]}}
{\deduce{\NSeq{\Sigma}{B_1[a_1/x], \ldots, B_n[a_n/x]}{B[(a_0~d).t/y, a_0/x]}}{r(x,\vec a,\Pi_2)}}
\enspace .
$$

\item Suppose $\Pi$ ends with $\eqL$:
$$
\infer[\eqL]
{\NSeq{\Sigma,x}{s = t, B_2, \ldots, B_n}{B_0}}
{
\left\{
\raisebox{-1.5ex}
{\deduce{\NSeq{(\Sigma,x)\theta}{B_2\theta,\ldots,B_n\theta}{B_0\theta}}{\Pi_\theta}}
\right\}_\theta
}
$$
where each $\theta$ is a unifier of $(\lambda \vec c.s, \lambda \vec c.t)$ and 
$\{ \vec c \} = supp(s = t).$
We need to show that for each unifier of 
$(\lambda a_1\lambda \vec c.s[a_1/x], \lambda a_1\lambda \vec c.t[a_1/x])$ 
there is a corresponding unifier for $\lambda \vec c.s$ and $\lambda \vec c.t.$ 
We can assume without loss of generality
that $x$ is not in the domain of $\rho$.

We first show the case where $x$ is not free in $\rho$. It is clear that
in this case $\rho$ is a unifier of $\lambda \vec c.s$ and $\lambda \vec c.t$. 
Therefore we apply the procedure recursively
to the premise derivation $\Pi_\rho$, to get the derivation
$r(x,\vec a,\Pi_\rho)$
of 
$$
\NSeq{\Sigma\rho}{(B_2[a_2/x])\rho, \ldots, (B_n[a_n/x])\rho}{(B_0[a_0/x])\rho}.
$$
In the other case, where $x$ is free in the range of $\rho$, 
we show that it can be reduced to the previous case.
First we define a substitution $\rho'$ to be the substitution $\rho$  where $x$
is replaced by a new variable $u$ which is not free in $\rho$.
Clearly $\rho'$ is also a unifier of 
$\lambda a_1\lambda \vec c.s[a_1/x]$ and $\lambda a_1\lambda \vec c.t[a_1/x].$
Moreover, it is more general than $\rho$, since $\rho = [x/u] \circ \rho'.$
Therefore we can apply the construction in the previous case
to get a derivation $r(x,\vec a,\Pi_{\rho'})$ and apply the substitution $[x/u]$ to 
to this derivation, using the procedure in Definition~\ref{def:subst}, 
to get a derivation of 
$$
\NSeq{\Sigma\rho}{(B_2[a_2/x])\rho, \ldots, (B_n[a_n/x])\rho}{(B_0[a_0/x])\rho.}
$$
The derivation $r(x,\vec a,\Pi)$ is then constructed as follows
$$
\infer[\eqL]
{\NSeq{\Sigma}{s[a_1/x] = t[a_1/x], \ldots, B_n[a_n/x]}{B_0[a_0/x]}}
{
\left\{
\raisebox{-1.5ex}{
\deduce{\NSeq{\Sigma\rho}{(B_2[a_2/x])\rho, \ldots, (B_n[a_n/x])\rho}{(B_0[a_0/x])\rho}}
  {\Pi_\rho'}
}
\right\}_{\rho}
}
$$
where each $\Pi_{\rho}'$ is constructed as explained above.

\item If $\Pi$ ends with $\cL$:
$$
\infer[\cL]
{\Seq{B_1,\ldots,B_j, \ldots, B_n}{B_0}}
{
\deduce{
\Seq{B_1,\ldots,B_j, B_j, \ldots, B_n}{B_0}
}
{\Pi'}
}
$$
then $r(x, \vec a, \Pi)$ is
$$
\infer[\cL]
{\Seq{B_1[a_1/x],\ldots,B_j[a_j/x], \ldots, B_n[a_n/x]}{B_0[a_0/x]}}
{
\deduce{
\Seq{B_1[a_1/x],\ldots,B_j[a_j/x], B_j[a_j/x], \ldots, B_n[a_n/x]}{B_0[a_0/x]}
}
{r(x, (a_0,\dots,a_j,a_j,\dots,a_n), \Pi')}
}
$$

\item If $\Pi$ ends with any other rule and has premise derivations $\Pi_1$,$\dots$, $\Pi_n$,
then $r(x,\vec a,\Pi)$ ends with the same rule and has premise derivations
$r(x,\vec a,\Pi_1)$, $\dots$, $r(x,\vec a,\Pi_n).$

\end{enumerate}
\end{definition}

\begin{lemma}
\label{lm:weak drv}
For any derivation $\Pi$ of $\NSeq{\Sigma}{\Gamma}{C}$ and any
multiset of $\Sigma$-formulas $\Delta$, $w(\Delta,\Pi)$ is
a derivation of $\NSeq{\Sigma}{\Gamma,\Delta}{C}$
and $ht(w(\Delta,\Pi)) \leq ht(\Pi).$
\end{lemma}

\begin{lemma}
\label{lm:subst drv}
For any derivation $\Pi$ of $\NSeq{\Sigma}{\Gamma}{C}$
and any $\Sigma$-substitution $\theta$, $\Pi\theta$ is
a derivation of $\NSeq{\Sigma\theta}{\Gamma\theta}{C\theta}$
and $ht(\Pi\theta) \leq ht(\Pi).$
\end{lemma}

\begin{lemma}
\label{lm:perm drv}
For any derivation $\Pi$ of $\Seq{B_1,\ldots,B_n}{B_0}$
and permutations $\vec \pi = \pi_0,\dots,\pi_n$,
$\langle \vec \pi \rangle.\Pi$ is a derivation of
$\Seq{\pi_1.B_1,\ldots,\pi_n.B_n}{\pi_0.B_0}$
and $ht(\langle \vec \pi\rangle.\Pi) \leq ht(\Pi).$
\end{lemma}

\begin{lemma}
\label{lm:res drv}
For any derivation $\Pi$ of $\NSeq{\Sigma,x}{B_1,\ldots,B_n}{B_0}$
and any list of nominal constants $\vec a = a_0,\dots,a_n$ such that
$a_i \not \in supp(B_i),$  $r(x,\vec a,\Pi)$ is a derivation
of $\NSeq{\Sigma}{B_1[a_1/x],\ldots,B_n[a_n/x]}{B_0[a_0/x]}$
and $ht(r(x,\vec a,\Pi)) \leq ht(\Pi).$
\end{lemma}

\begin{lemma}
\label{lm:subst}{\em Substitutions.} 
Let $\Pi$ be a proof of $\NSeq{\Sigma}{\Gamma}{C}$
and let $\theta$ be a $\Sigma$-substitution.
Then there exists a proof $\Pi'$ of
$\NSeq{\Sigma\theta}{\Gamma\theta}{C\theta}$ such
that $ht(\Pi') \leq ht(\Pi).$
\end{lemma}
\begin{proof}
Follows immediately from Lemma~\ref{lm:subst drv}.
\qed
\end{proof}

\begin{lemma}
\label{lm:perm} {\em Permutations.}
Let $\Pi$ be a proof of $\NSeq \Sigma {B_1, \ldots, B_n} {B_0}.$
Then there exists a proof $\Pi'$ of 
$\NSeq{\Sigma}{\pi_1.B_1, \ldots, \pi_n.B_n}{\pi_0.B_0}$
such that $ht(\Pi') \leq ht(\Pi).$
\end{lemma}
\begin{proof}
Follows immediately from Lemma~\ref{lm:perm drv}.
\qed
\end{proof}

\begin{lemma}
\label{lm:res}{\em Restricted name substitutions.}
Let $\Pi$ be a proof of 
$$\NSeq{\Sigma,x:\iota}{B_1, \ldots, B_n}{B_0}.$$
Then there exists a proof of $\Pi'$ of
$\NSeq{\Sigma}{B_1[a_1/x], \ldots, B_n[a_n/x]}{B_0[a_0/x]},$
where $a_i \not \in supp(B_i)$ for each $i \in \{0,\dots,n\},$
such that $ht(\Pi') \leq ht(\Pi).$
\end{lemma}
\begin{proof}
Follows immediately from Lemma~\ref{lm:res drv}.
\qed
\end{proof}

The next two lemmas are crucial to the cut-elimination proof: they allow
one to reintroduce the symmetry between $\forallL$ and $\forallR$,
and dually, between $\existsL$ and $\existsR$ rules.

\begin{lemma}
\label{lm:supp1}{\em Support extension.}
Let $\Pi$ be a proof of $\NSeq{\Sigma,h}{\Gamma}{B[h~\vec{a}/x]}$
where $\{\vec a\} = supp(B)$, 
$h \not \in \Sigma$ and $h$ is not free in $\Gamma$ and $B$. 
Let $\vec{c}$ be a list of nominal constants not in the support
of $B$. Then there exists a proof $\Pi'$ of
$\NSeq{\Sigma, h'}{\Gamma}{B[h'~\vec{a}\vec{c}/x]}$ where
$h' \not \in \Sigma.$
\end{lemma}
\begin{proof}
Suppose $\vec c$ is the list of constants 
$c_1 : \iota_1,\ldots,c_n : \iota_n$. 
Let $\vec y = y_1 : \iota_1,\ldots, y_n:\iota_n$
be a list of distinct variables not appearing in $\Sigma \cup \{h, h'\}$. 
We first apply the substitution $[\lambda \vec a.h'\,\vec c \vec x/h]$
to the sequent $\NSeq{\Sigma,h}{\Gamma}{B[h\vec a/x]}.$
By Lemma~\ref{lm:subst}, there is a proof $\Pi_1$ of
$$
\NSeq{\Sigma, h', \vec y}{\Gamma}{B[h'\,\vec a \vec y/x]}
$$
The derivation $\Pi'$ is then obtained by repeatedly applying 
Lemma~\ref{lm:res} to $\Pi_1$ to change $\vec y$ into
$\vec c$.
\qed
\end{proof}

\begin{lemma}
\label{lm:supp2}{\em Support extension.}
Let $\Pi$ be a proof of $\NSeq{\Sigma,h}{B[h~\vec{a}/x], \Gamma}{C}$
where $\{\vec a\} = supp(B)$, 
$h \not \in \Sigma$ and $h$ is not free in $\Gamma$, $B$ and $C$. 
Let $\vec{c}$ be a list of nominal constants not in the support
of $B$. Then there exists a proof $\Pi'$ of
$\NSeq{\Sigma, h'}{B[h'~\vec{a}\vec{c}/x], \Gamma}{C}$ where
$h' \not \in \Sigma.$
\end{lemma}
\begin{proof}
Use the same construction as in the proof of Lemma~\ref{lm:supp1}.
\qed
\end{proof}

\section{Cut reduction}
\label{sec:reduct}

We define a {\em reduction} relation between derivations, following closely
the reduction relation in ~\cite{mcdowell00tcs}.
For simplicity of presentation, we shall omit the signatures in the sequents
in the following reduction of cuts when the signatures are not changed 
by the reduction or when it is clear from context
which signatures should be assigned to the sequents.
The redex is always a derivation $\Xi$ ending with the multicut rule
\begin{displaymath}
\infer[\mc]{\NSeq{\Sigma}{\Delta_1,\ldots,\Delta_n,\Gamma}{C}}
        {\deduce{\NSeq{\Sigma}{\Delta_1}{B_1}}
                {\Pi_1}
        & \cdots
        & \deduce{\NSeq{\Sigma}{\Delta_n}{B_n}}
                {\Pi_n}
        & \deduce{\NSeq{\Sigma}{B_1,\ldots,B_n,\Gamma}{C}}
                {\Pi}}
\enspace .
\end{displaymath}
We refer to the formulas $B_1,\dots,B_n$ produced by the
$\mc$ as {\em cut formulas}.

If $n=0$, $\Xi$ reduces to the premise derivation $\Pi$.

For $n > 0$ we specify the reduction relation based on the last rule
of the premise derivations.
If the rightmost premise derivation $\Pi$ ends with a left rule acting
on a cut formula $B_i$, then the last rule of $\Pi_i$ and the last
rule of $\Pi$ together determine the reduction rules that apply.
We classify these rules according to the following criteria: we call
the rule an {\em essential} case when $\Pi_i$ ends with a right rule;
if it ends with a left rule, it is a {\em left-commutative} case;
if $\Pi_i$ ends with the $\init$ rule, then we have an {\em axiom}
case; a {\em multicut} case arises when it ends with the $\mc$ rule.
When $\Pi$ does not end with a left rule acting on a cut formula, then
its last rule is alone sufficient to determine the reduction rules
that apply.
If $\Pi$ ends in a rule acting on a formula other than a cut
formula, then we call this a {\em right-commutative} case.
A {\em structural} case results when $\Pi$ ends with a contraction or weakening on
a cut formula.
If $\Pi$ ends with the $\init$ rule, this is also an axiom case;
similarly a multicut case arises if $\Pi$ ends in the $\mc$ rule.

For simplicity of presentation, we always show $i = 1$.

\noindent{\em \underline{Essential cases:}}

\noindent $\landR/\landL$: If $\Pi_1$ and $\Pi$ are
\begin{displaymath}
\infer[\landR]{\Seq{\Delta_1}{B_1' \land B_1''}}
        {\deduce{\Seq{\Delta_1}{B_1'}}
                {\Pi_1'}
        & \deduce{\Seq{\Delta_1}{B_1''}}
                {\Pi_1''}}
\qquad\qquad\qquad
\infer[\landL]{\Seq{B_1' \land B_1'',B_2,\ldots,B_n,\Gamma}{C}}
        {\deduce{\Seq{B_1',B_2,\ldots,B_n,\Gamma}{C}}
                {\Pi'}}
\enspace ,
\end{displaymath}
then $\Xi$ reduces to
\begin{displaymath}
\infer[\mc]{\Seq{\Delta_1,\ldots,\Delta_n,\Gamma}{C}}
        {\deduce{\Seq{\Delta_1}{B_1'}}
                {\Pi_1'}
        & \deduce{\Seq{\Delta_2}{B_2}}
                {\Pi_2}
        & \cdots
        & \deduce{\Seq{\Delta_n}{B_n}}
                {\Pi_n}
        & \deduce{\Seq{B_1',B_2,\ldots,B_n,\Gamma}{C}}
                {\Pi'}}
\enspace .
\end{displaymath}
The case for the other $\landL$ rule is symmetric.

\vskip12pt

\noindent $\lorR/\lorL$: If $\Pi_1$ and $\Pi$ are
\begin{displaymath}
\infer[\lorR]{\Seq{\Delta_1}{B_1' \lor B_1''}}
        {\deduce{\Seq{\Delta_1}{B_1'}}
                {\Pi_1'}}
\qquad\qquad\!\!\!
\infer[\lorL]{\Seq{B_1' \lor B_1'',B_2,\ldots,B_n,\Gamma}{C}}
        {\deduce{\Seq{B_1',B_2,\ldots,B_n,\Gamma}{C}}
                {\Pi'}
        & \deduce{\Seq{B_1'',B_2,\ldots,B_n,\Gamma}{C}}
                {\Pi''}}
\enspace ,
\end{displaymath}
then $\Xi$ reduces to
\begin{displaymath}
\infer[\mc]{\Seq{\Delta_1,\ldots,\Delta_n,\Gamma}{C}}
        {\deduce{\Seq{\Delta_1}{B_1'}}
                {\Pi_1'}
        & \deduce{\Seq{\Delta_2}{B_2}}
                {\Pi_2}
        & \cdots
        & \deduce{\Seq{\Delta_n}{B_n}}
                {\Pi_n}
        & \deduce{\Seq{B_1',B_2,\ldots,B_n,\Gamma}{C}}
                {\Pi'}}
\enspace .
\end{displaymath}
The case for the other $\lorR$ rule is symmetric.

\vskip12pt

\noindent $\oimpR/\oimpL$: Suppose $\Pi_1$ and $\Pi$ are
\begin{displaymath}
\infer[\oimpR]{\Seq{\Delta_1}{B_1' \oimp B_1''}}
        {\deduce{\Seq{B_1',\Delta_1}{B_1''}}
                {\Pi_1'}}
\qquad\qquad\!\!
\infer[\oimpL]{\Seq{B_1' \oimp B_1'',B_2,\ldots,B_n,\Gamma}{C}}
        {\deduce{\Seq{B_2,\ldots,B_n,\Gamma}{B_1'}}
                {\Pi'}
        & \deduce{\Seq{B_1'',B_2,\ldots,B_n,\Gamma}{C}}
                {\Pi''}}
\enspace .
\end{displaymath}
Let $\Xi_1$ be
\begin{displaymath}
\infer[\mc]{\Seq{\Delta_1,\ldots,\Delta_n,\Gamma}{B_1''}}
        {\infer[\mc]{\Seq{\Delta_2,\ldots,\Delta_n,\Gamma}{B_1'}}
                {\left\{\raisebox{-1.5ex}{\deduce{\Seq{\Delta_i}{B_i}}
                        {\Pi_i}}\right\}_{i \in \{2..n\}}
                & \raisebox{-2.5ex}{\deduce{\Seq{B_2,\ldots,B_n,\Gamma}{B_1'}}
                        {\Pi'}}}
        & \deduce{\Seq{B_1',\Delta_1}{B_1''}}
                {\Pi_1'}}
\enspace .
\end{displaymath}
Then $\Xi$ reduces to
\settowidth{\infwidthi}
        {$\Seq{\Delta_1,\ldots,\Delta_n,\Gamma,\Delta_2,\ldots,\Delta_n,\Gamma}{C}$}
\begin{displaymath}
\infer{\Seq{\Delta_1,\ldots,\Delta_n,\Gamma}{C}}
        {\infer[\cL]{\makebox[\infwidthi]{}}
                {\infer[\mc]{\Seq{\Delta_1,\ldots,\Delta_n,\Gamma,
                                        \Delta_2,\ldots,\Delta_n,\Gamma}{C}}
                        {\raisebox{-2.5ex}{\deduce{\Seq{\ldots}{B_1''}}
                                {\Xi_1}}
                        & \left\{\raisebox{-1.5ex}{\deduce{\Seq{\Delta_i}{B_i}}
                                {\Pi_i}}\right\}_{i \in \{2..n\}}
                        & \raisebox{-2.5ex}{\deduce{\Seq{B_1'',\{B_i\}_{i \in \{2..n\}},\Gamma}{C}}
                                {\Pi''}}}}}
\enspace .
\end{displaymath}
We use the double horizontal lines to indicate that the relevant
inference rule (in this case, $\cL$) may need to be applied zero or
more times.

\vskip12pt

\noindent $\forallR/\forallL$: Suppose $\Pi_1$ and $\Pi$ are
\begin{displaymath}
\infer[\forallR]
{\NSeq{\Sigma}{\Delta_1}{\forall x.B_1'}}
{\deduce{\NSeq{\Sigma, h}{\Delta_1}{B_1'[(h\,\vec{c})/x]}}
                {\Pi_1'}}
\qquad\qquad\qquad
\infer[\forallL]
{\NSeq{\Sigma}{\forall x.B_1',B_2,\ldots,B_n,\Gamma}{C}}
{\deduce{\NSeq{\Sigma}{B_1'[t/x],B_2,\ldots,B_n,\Gamma}{C}}
                {\Pi'}}
\enspace ,
\end{displaymath}
where $\{\vec c\} = supp(B_1').$
Let $\{\vec d\} = supp(B_1'[t/x]) \setminus supp(B_1').$
Apply Lemma~\ref{lm:supp1} to get a derivation
$\Pi_1''$ of $\NSeq{\Sigma,h'}{\Delta_1}{B_1'[(h\,\vec c\vec d)/x]}.$
The derivation $\Xi$ reduces to

\begin{displaymath}
\infer[\mc]
{\NSeq{\Sigma}{\Delta_1,\ldots,\Delta_n,\Gamma}{C}}
{\raisebox{-2.5ex}
   {\deduce{\NSeq{\Sigma}{\Delta_1}{B_1'[t/x]}}
           {\Pi_1''[\lambda\vec{c}\vec d.t/h']}}
  & \left\{\raisebox{-1.5ex}
           {\deduce{\NSeq{\Sigma}{\Delta_i}{B_i}}
                   {\Pi_i}}\right\}_{i \in \{2..n\}}
  & \raisebox{-2.5ex}{\deduce{\Seq{\ldots}{C}}
                {\Pi'}}}
\enspace .
\end{displaymath}

\vskip12pt

\noindent $\existsR/\existsL$:
Suppose $\Pi_1$ and $\Pi$ are
\begin{displaymath}
\infer[\existsR]
{\NSeq{\Sigma}{\Delta_1}{\exists x.B_1'}}
        {\deduce{\NSeq{\Sigma}{\Delta_1}{B_1'[t/x]}}
                {\Pi_1'}}
\qquad\qquad\qquad
\infer[\existsL]
{\NSeq{\Sigma}{\exists x.B_1',B_2,\ldots, B_n,\Gamma}{C}}
{\deduce{\NSeq{\Sigma,h}{B_1'[(h\,\vec{c})/x],B_2,\ldots,B_n,
            \Gamma}{C}}
                {\Pi'}}
\enspace ,
\end{displaymath}
where $\{\vec c\} = supp(B_1').$
Let $\{\vec d\} = supp(B_1'[t/x])\setminus supp(B_1').$
Apply Lemma~\ref{lm:supp2} to $\Pi'$ to get a derivation $\Pi''$
of $\NSeq{\Sigma,h'}{\Delta_1}{B_1'[(h'\,\vec c\vec d)/x]}.$
Then $\Xi$ reduces to
\begin{displaymath}
\infer[\mc]{\NSeq{\Sigma}{\Delta_1,\ldots,\Delta_n,\Gamma}{C}}
  {\deduce{\NSeq{\Sigma}{\Delta_1}{B_1'[t/x]}}{\Pi_1'}
  & \ldots
  & \deduce{\NSeq{\Sigma}{B_1'[t/x], B_2,\dots,\Gamma}{C}}
              {\Pi''[\lambda \vec{c}\vec d.t/h']}
        }
\enspace .
\end{displaymath}

\vskip12pt

\noindent $\nablaR/\nablaL$: Suppose $\Pi_1$ and $\Pi$ are
\begin{displaymath}
\infer[\nablaR]
{\Seq{\Delta_1}{\nabla x.B_1'}}
     {\deduce{\Seq{\Delta_1}{B_1'[a/x]}}
           {\Pi_1'}}
\qquad\qquad\qquad
\infer[\nablaL]{\Seq{\nabla x.B_1',\ldots,
       B_n,\Gamma}{C}}
  {\deduce{\Seq{B_1'[b/x],\ldots,B_n,
            \Gamma}{C}}
                {\Pi'}}
\enspace .
\end{displaymath}
Apply the construction in Definition~\ref{def:perm}
to to $\Pi_1'$ to swap $a$ with $b$ to get a derivation $\Pi_1''$
of $\Seq{\Delta_1}{B_1'[b/x]}.$
$\Xi$ reduces to
\begin{displaymath}
\infer[\mc]
{\Seq{\Delta_1,\ldots,\Delta_n,\Gamma}{C}}
{
   \deduce{\Seq{\Delta_1}{B_1'[b/x]}}{\Pi_1''}
    & \ldots
        & 
    \deduce{\Seq{B_1'[b/x], \dots,B_n, \Gamma}{C}}
           {\Pi'}
}
\enspace .
\end{displaymath}

\vskip12pt

\noindent $\natR/\natL:$
Suppose $\Pi_1$ is
$
\infer[\natR]
{\Seq {\Delta_1} {nat~z}}
{}
$
and $\Pi$ is
$$
\infer[\natL]
{\Seq{nat~z, B_2, \ldots, B_n, \Gamma}{C}}
{
\deduce{\Seq{}{D~z}}{\Pi'} &
\deduce{\Seq{D~j}{D~(s\,j)}}{\Pi''} &
\deduce{\Seq{D~z, B_2, \ldots, B_n, \Gamma}{C}}{\Pi'''}
}
\enspace .
$$
Then $\Xi$ reduces to
\begin{displaymath}
   \infer[\mc]
   {\Seq{\Delta_1,\Delta_2,\ldots,\Delta_n,\Gamma}{C}}
   {\raisebox{-1.5ex}{\deduce{\Seq {\Delta_1}{D~z}}{w(\Delta_1,\Pi')}} &
    \left\{\raisebox{-1.5ex}
      {\deduce{\Seq {\Delta_i}{B_i}}{\Pi_i}}
    \right\}_{i\in\{2\dots n\}} &
    \raisebox{-1.5ex}{\deduce{\Seq{D~z,  B_2, \ldots, B_n, \Gamma}{C}}{\Pi'''}}
   }
\end{displaymath}

\vskip12pt

\noindent $\natR/\natL:$
Suppose $\Pi_1$ is
$$
\infer[\natR]
{\Seq{\Delta_1}{nat~(s\,I)}}
{\deduce{\Seq{\Delta}{nat\,I}}{\Pi_1'}}
$$
and $\Pi$ is
$$
\infer[\natL]
{\Seq{nat~(s\,I), B_2, \ldots, B_n, \Gamma} C}
{
\deduce{\Seq{}{D\,z}}{\Pi'} &
\deduce{\Seq{D\,j}{D\,(s\,j)}}{\Pi''} &
\deduce{\Seq{D\,(s\,I), B_2, \ldots, B_n, \Gamma}{C}}{\Pi'''}
}
$$
Let $\Xi_1$ be
$$
\infer[mc.]
{\Seq{\Delta_1}{D~I}}
{
\deduce{\Seq{\Delta_1}{nat\,I}}{\Pi_1'}
&
\infer[\natL]
{\Seq{nat\,I}{D\,I}}
{
\deduce{\Seq{}{D\,z}}{\Pi'}
&
\deduce{\Seq{D\,j}{D\,(s\,j)}}{\Pi''}
&
\infer[id_\pi]{\Seq{D\,I}{D\,I}}{}
}
}
$$
Suppose $\{\vec c \} = supp(I).$
We apply the procedures in Definition~\ref{def:subst}
and Definition~\ref{def:res} to $\Pi''$ to obtain
the derivation $\Pi^\bullet$ of
$$
\NSeq{h}{D\,(h\,\vec c)}{D\,(s\,(h\,\vec c))}.
$$
Let $\Xi_2$ be
$$
\infer[mc.]
{\Seq{\Delta_1}{D\,(s\,I)}}
{
\deduce{\Seq{\Delta_1}{D\,I}}{\Xi_1}
&
\deduce{\Seq{D\,I}{D\,(s\,I)}}{\Pi^{\bullet}[\lambda \vec c.I/h]}
}
$$
Then $\Xi$ reduces to
$$
\infer[mc.]
{\Seq{\Delta_1,\ldots,\Delta_n,\Gamma}{C}}
{
\deduce{\Seq{\Delta_1}{D\,(s\,I)}}{\Xi_2}
&
\deduce{\Seq{\Delta_2}{B_2}}{\Pi_2}
&
\ldots
&
\deduce{\Seq{\Delta_n}{B_n}}{\Pi_n}
&
\deduce{\Seq{D\,(s\,I), B_2,\dots,B_n,\Gamma}{C}}{\Pi'''}
}
$$

\vskip12pt

\noindent $\eqL/\eqR:$
If $\Pi_1$ and $\Pi$ are
$$
\infer[\eqR]
{\NSeq{\Sigma}{\Delta_1}{t=t}}
{}
\qquad
\infer[\eqL]
{\NSeq{\Sigma}{t=t, \Gamma}{C}}
{
\left\{
\raisebox{-1.5ex}
{\deduce{\NSeq{\Sigma\theta}{\Gamma\theta}{C\theta}}{\Pi_\theta}}
\right\}_\theta
}
$$
then $\Xi$ reduces to 
$$
\infer[mc]
{\NSeq{\Sigma}{\Delta_1,\ldots,\Delta_n,\Gamma}{C}}
{
\deduce{\NSeq{\Sigma}{\Delta_2}{B_2}}{\Pi_2}
&
\ldots
&
\deduce{\NSeq{\Sigma}{\Delta_n}{B_n}}{\Pi_n}
&
\deduce{\NSeq{\Sigma}{\Delta_1,B_2,\ldots,B_n,\Gamma}{C}}{w(\Delta_1,\Pi_\epsilon)}
}
$$
where $\epsilon$ is the empty substitution.

\noindent $\defR/\defL$:
Suppose $\Pi_1$ and $\Pi$ are
$$
\infer[\defR]
{\Seq{\Delta_1}{p\,\bar{t}}}
  {\deduce{\Seq{\Delta_1}{B[\vec{t}/\vec x]}}{\Pi_1'}}
\qquad \qquad
\infer[\defL]{\Seq{p\,\vec{t},B_2,\dots,\Gamma}{C}}
  {
   \deduce
      {\Seq{B[\vec{t}/\vec x],B_2, \dots,\Gamma}{C}}
      {\Pi'}
  }  
\enspace .  
$$
Then $\Xi$ reduces to
$$
\infer[\mc]
{\Seq{\Delta_1,\dots,\Delta_n,\Gamma}{C}}
{
\deduce{\Seq{\Delta_1}{B[\vec{t}/\vec x]}}
   {\Pi_1'}
&
\deduce{\Seq{\Delta_2}{B_2}}{\Pi_2}
&
\ldots
&
\deduce{\Seq{\Delta_n}{B_n}}{\Pi_n}
&
\deduce{\Seq{B[\vec t/\vec x], \dots, \Gamma}{C}}{\Pi'}   
}
\enspace .
$$

\vskip12pt

\newcommand{\bulletL}{\bullet{\cal L}}
\newcommand{\circL}{\circ{\cal L}}

\noindent{\em \underline{Left-commutative cases:}}

\vskip12pt

\noindent $\bulletL/\circL$:
Suppose $\Pi$ ends with a left rule other than $\cL$
acting on $B_1$ and $\Pi_1$ is
\begin{displaymath}
\infer[\bulletL]
{\Seq{\Delta_1}{B_1}}
{\left\{
  \raisebox{-1.5ex}
  {\deduce{\Seq{\Delta_1^i}{B_1}
  }
  {\Pi_1^i}}
\right\}
}
\enspace ,
\end{displaymath}
where $\bulletL$ is any left rule except $\oimpL$, $\eqL$, or
$\natL$. 
Then $\Xi$ reduces to
\settowidth{\infwidthi}
{$\left\{\raisebox{-3.5ex}
  {\infer[\mc]{\Seq{\Delta_1^i,\Delta_2,\ldots,\Delta_n,\Gamma}
    {C}}
  {\raisebox{-2.5ex}{\deduce{\Seq{\Delta_1^i}{B_1}}
       {\Pi_1^i}}
  & \left\{\raisebox{-1.5ex}{\deduce{\Seq{\Delta_j}{B_j}}
                        {\Pi_j}}\right\}_{j \in \{2..n\}}
                & \raisebox{-2.5ex}{\deduce{\Seq{B_1,\ldots,B_n,\Gamma}{C}}
                        {\Pi}}}}\right\}$}
\settowidth{\infwidthii}{\mc}
\begin{displaymath}
\infer[\bulletL]
{\Seq{\Delta_1,\Delta_2,\ldots,\Delta_n,\Gamma}{C}}
{\deduce{\makebox[\infwidthi]{}}
  {\left\{\raisebox{-3.5ex}
      {\infer[\mc]
          {\Seq{\Delta_1^i,\Delta_2,\ldots,\Delta_n,\Gamma}{C}}
          {\raisebox{-2.5ex}{
             \deduce{\Seq{\Delta_1^i}{B_1}}
                    {\Pi_1^i}}
         & \left\{\raisebox{-1.5ex}{\deduce{\Seq{\Delta_j}{B_j}}
                 {\Pi_j}}\right\}_{j \in \{2..n\}}
         & \raisebox{-2.5ex}{\deduce{\Seq{B_1,\ldots,B_n,\Gamma}{C}}
                 {\Pi}}}}\right\}\makebox[\infwidthii]{}}}
\enspace .
\end{displaymath}

\vskip12pt
\noindent $\oimpL/\circL$:
Suppose $\Pi$ ends with a left rule other than $\cL$ acting
on $B_1$ and $\Pi_1$ is
\begin{displaymath}
\infer[\oimpL]
{\Seq{D_1' \oimp D_1'',\Delta_1'}{B_1}}
{\deduce{\Seq{\Delta_1'}{D_1'}}
                {\Pi_1'}
        & \deduce{\Seq{D_1'',\Delta_1'}{B_1}}
                {\Pi_1''}}
\enspace .
\end{displaymath}
Let $\Xi_1$ be
\begin{displaymath}
\infer[\mc]
{\Seq{D_1'',\Delta_1',\Delta_2,\ldots,\Delta_n,\Gamma}{C}}
        {\deduce{\Seq{D_1'',\Delta_1'}{B_1}}
                {\Pi_1''}
        & \deduce{\Seq{\Delta_2}{B_2}}
                {\Pi_2}
        & \cdots
        & \deduce{\Seq{\Delta_n}{B_n}}
                {\Pi_n}
        & \deduce{\Seq{B_1,\ldots,B_n,\Gamma}{C}}
                {\Pi}}
\enspace .
\end{displaymath}
Then $\Xi$ reduces to
\begin{displaymath}
\infer[\oimpL]
{\Seq{D_1' \oimp D_1'',\Delta_1',\Delta_2,\ldots,\Delta_n,\Gamma}{C}}
{
 \deduce{\Seq{\Delta_1',\Delta_2,\ldots,\Delta_n,\Gamma}{D_1'}}
  {w(\Delta_2\cup \dots \cup \Delta_n\cup \Gamma, \Pi_1')}
  & 
\deduce{\Seq{D_1'',\Delta_1',\Delta_2,\ldots,\Delta_n,\Gamma}{C}}
                {\Xi_1}
}
\enspace .
\end{displaymath}

\vskip12pt

\noindent $\natL/\circL:$
Suppose $\Pi$ ends with a left rule other than $\cL$ acting on $B_1$
and $\Pi_1$ is
$$
\infer[\natL]
{\Seq{nat\,I, \Delta_1'}{B_1}}
{\deduce{\Seq{}{D_1\,z}}{\Pi_1^1}
& \deduce{\Seq{D_1\,j}{D_1(s\,j)}}{\Pi_1^2} 
& \deduce{\Seq{D_1 I, \Delta_1'}{B_1}}{\Pi_1^3}
}
$$
Let $\Xi_1$ be
$$
\infer[mc]
{\Seq{D_1 I, \Delta_1', \Delta_2, \ldots, \Delta_n, \Gamma}{C}}
{
\deduce{\Seq{D_1 I, \Delta_1'}{B_1}}{\Pi_1^3}
&
\deduce{\Seq{\Delta_2}{B_2}}{\Pi_2}
&
\ldots
&
\deduce{\Seq{\Delta_n}{B_n}}{\Pi_n}
&
\deduce{\Seq{B_1,\ldots,B_n}{C}}{\Pi}
}
$$
Then $\Xi$ reduces to
$$
\infer[\natL]
{\Seq{nat\,I, \Delta_1', \Delta_2, \ldots,\Delta_n,\Gamma}{C}}
{
\deduce{\Seq{}{D_1 z}}{\Pi_1^1}
&
\deduce{\Seq{D_1 j}{D_1 (s\,j)}}{\Pi_1^2}
&
\deduce{\Seq{D_1 I, \Delta_1', \Delta_2,\ldots,\Delta_2,\Gamma}{C}}{\Xi_1}
}
$$

\vskip12pt

\noindent $\eqL/\circL:$
If $\Pi$ ends with a left rule other than $\cL$ acting
on $B_1$ and $\Pi_1$ is
$$
\infer[\eqL]
{\Seq{s=t, \Delta_1'}{B_1}}
{
\left\{
\raisebox{-1.5ex}
{\deduce{\Seq{\Delta_1'\theta}{B_1\theta}}{\Pi^\theta}}
\right\}_\theta
}
$$
then $\Xi$ reduces to
$$
\infer[\eqL]
{
\Seq{\qquad \qquad \qquad \quad
s=t,\Delta_1',\Delta_2,\ldots,\Delta_n,\Gamma}{C
\qquad \qquad \qquad \qquad \qquad \qquad}
}
{
\left\{
\raisebox{-1.5ex}
{
\infer[mc]
{\Seq{\Delta_1'\theta, \Delta_2\theta,\ldots, \Delta_n\theta,
\Gamma\theta}{C\theta}}
{
\deduce{\Seq{\Delta_1'\theta}{B_1\theta}}{\Pi^\theta}
&
\deduce{\Seq{\Delta_2\theta}{B_2\theta}}{\Pi_2\theta}
&
\ldots
&
\deduce{\Seq{\Delta_n\theta}{B_n\theta}}{\Pi_n\theta}
&
\deduce{\Seq{B_1\theta,\ldots,B_n\theta,\Gamma\theta}{C\theta}}{\Pi\theta}
}
}
\right\}_\theta
}
$$

\vskip12pt

\noindent{\em \underline{Right-commutative cases:}}

\vskip12pt

\noindent $-/\circL$:
Suppose $\Pi$ is
\begin{displaymath}
\infer[\circL]
{\Seq{B_1,\ldots,B_n,\Gamma}{C}}
{\left\{\raisebox{-1.5ex}
{\deduce{\Seq{B_1,\ldots,B_n,\Gamma^i}{C}}
 {\Pi^i}}\right\}}
\enspace ,
\end{displaymath}
where $\circL$ is any left rule other than $\oimpL$, $\eqL$,
or $\natL$ (but including $\cL$) acting on a formula other
than $B_1, \ldots, B_n$. 
The derivation $\Xi$ reduces to
\settowidth{\infwidthi}
        {$\left\{\raisebox{-2.45ex}{
          \infer[\mc]{\Seq{\Delta_1,\ldots,\Delta_n,\Gamma^i}{C}}
                {\deduce{\Seq{\Delta_1}{B_1}}
                        {\Pi_1'}
                & \cdots
                & \deduce{\Seq{\Delta_n}{B_n}}
                        {\Pi_n'}
                & \deduce{\Seq{B_1,\ldots,B_n,\Gamma^i}{C}}
                        {\Pi^i}}}\right\}$}
\settowidth{\infwidthii}{\mc}
\begin{displaymath}
\infer[\circL]{\Seq{\Delta_1,\ldots,\Delta_n,\Gamma}{C}}
        {\deduce{\makebox[\infwidthi]{}}
            {\left\{\raisebox{-2.45ex}{
             \infer[\mc]{\Seq{\Delta_1,\ldots,\Delta_n,\Gamma^i}{C}}              
             {\deduce{\Seq{\Delta_1}{B_1}}
                        {\Pi_1}
                & \cdots
                & \deduce{\Seq{\Delta_n}{B_n}}
                        {\Pi_n}
                & \deduce{\Seq{B_1,\ldots,B_n,\Gamma^i}{C}}
                        {\Pi^i}}}\right\}\makebox[\infwidthii]{}}}
\enspace ,
\end{displaymath}

\noindent $-/\oimpL$:
Suppose $\Pi$ is
\begin{displaymath}
\infer[\oimpL]{\Seq{B_1,\ldots,B_n, D' \oimp D'',\Gamma'}{C}}
        {\deduce{\Seq{B_1,\ldots,B_n,\Gamma'}{D'}}
                {\Pi'}
        & \deduce{\Seq{B_1,\ldots,B_n,D'',\Gamma'}{C}}
                {\Pi''}}
\enspace .
\end{displaymath}
Let $\Xi_1$ be
\begin{displaymath}
\infer[\mc]{\Seq{\Delta_1,\ldots,\Delta_n,\Gamma'}{D'}}
        {\deduce{\Seq{\Delta_1}{B_1}}
                {\Pi_1}
        & \cdots
        & \deduce{\Seq{\Delta_n}{B_n}}
                {\Pi_n}
        & \deduce{\Seq{B_1,\ldots,B_n,\Gamma'}{D'}}
                {\Pi'}}
\end{displaymath}
and $\Xi_2$ be
\begin{displaymath}
\infer[\mc]{\Seq{\Delta_1,\ldots,\Delta_n,D'',\Gamma'}{C}}
        {\deduce{\Seq{\Delta_1}{B_1}}
                {\Pi_1}
        & \cdots
        & \deduce{\Seq{\Delta_n}{B_n}}
                {\Pi_n}
        & \deduce{\Seq{B_1,\ldots,B_n,D'',\Gamma'}{C}}
                {\Pi''}}
\enspace .
\end{displaymath}
Then $\Xi$ reduces to
\begin{displaymath}
\infer[\oimpL]{\Seq{\Delta_1,\ldots,\Delta_n,D' \oimp D'',\Gamma'}{C}}
        {\deduce{\Seq{\Delta_1,\ldots,\Delta_n,\Gamma'}{D'}}
                {\Xi_1}
        & \deduce{\Seq{\Delta_1,\ldots,\Delta_n,D'',\Gamma'}{C}}
                {\Xi_2}}
\enspace .
\end{displaymath}

\vskip12pt

\noindent $-/\natL:$
Suppose $\Pi$ is
$$
\infer[\natL]
{\Seq{B_1,\ldots,B_n, nat\,I, \Gamma'}{C}}
{
\deduce{\Seq{}{D\,z}}{\Pi'}
&
\deduce{\Seq{D\,j}{D\,(s\,j)}}{\Pi''}
&
\deduce{\Seq{B_1,\ldots,B_n,D\,I,\Gamma'}{C}}{\Pi'''}
}
$$
Let $\Xi_1$ be 
$$
\infer[mc,]
{\Seq{\Delta_1,\ldots,\Delta_n,D\,I,\Gamma'}{C}}
{
\deduce{\Seq{\Delta_1}{B_1}}{\Pi_1}
&
\ldots
&
\deduce{\Seq{\Delta_n}{B_n}}{\Pi_n}
&
\deduce{\Seq{B_1,\ldots,B_n,D\,I, \Gamma'}{C}}{\Pi'''}
}
$$
then $\Xi$ reduces to
$$
\infer[\natL]
{\Seq{\Delta_1,\ldots,\Delta_n,nat\,I,\Gamma'}{C}}
{
\deduce{\Seq{}{D\,z}}{\Pi'}
&
\deduce{\Seq{D\,j}{D\,(s\,j)}}{\Pi''}
&
\deduce{\Seq{\Delta_1,\ldots,\Delta_n,D\,I,\Gamma'}{C}}{\Xi_1}
}
$$

\vskip12pt

\noindent $-/\eqL$:
If $\Pi$ is
\begin{displaymath}
\infer[\eqL]{\Seq{B_1,\ldots,B_n,{s=t},
                 \Gamma'}{C}}
        {\left\{\raisebox{-1.5ex}
                {\deduce{\Seq{B_1\rho,\ldots,
                                 B_n\rho,\Gamma'\rho}
                                {C\rho}}
                {\Pi^{\rho}}}\right\}}
\enspace ,
\end{displaymath}
then $\Xi$ reduces to
\settowidth{\infwidthi}
        {$\left\{\raisebox{-3.5ex}{\infer[\mc]
                        {\Seq{\Delta_1\rho,\ldots,
                                 \Delta_n\rho,D\sigma,\Gamma'\rho}
                                {C\rho}}
                {\left\{\raisebox{-1.5ex}
                        {\deduce{\Seq{\Delta_i\rho}{B_i\rho}}
                                {\Pi_i\rho}}
                 \right\}_{i \in \{1..n\}}
                & \raisebox{-2.5ex}
                        {\deduce{\Seq{\{B_i\rho\}_{i\in\{1..n\}},
                                         D\sigma,\Gamma'\rho}
                                        {C\rho}}
                                {\Pi^{\rho,\sigma,D}}}}}
        \right\}$}
\settowidth{\infwidthii}{\mc}
\begin{displaymath}
\infer[\eqL]
{\Seq{\Delta_1,\ldots,\Delta_n,s=t,\Gamma'}{C}}
     {\deduce{\makebox[\infwidthi]{}}
     {\left\{\raisebox{-3.5ex}{\infer[\mc]
            {\Seq{\Delta_1\rho,\ldots,
                  \Delta_n\rho,\Gamma'\rho}
                                        {C\rho}}
       {\left\{\raisebox{-1.5ex}
          {\deduce{\Seq{\Delta_i\rho}{B_i\rho}}
                                {\Pi_i\rho}}
                 \right\}_{i \in \{1..n\}}
                & \raisebox{-2.5ex}
                        {\deduce{\Seq{B_i\rho, \ldots,
                                 \Gamma'\rho}
                                {C\rho}}
                        {\Pi^{\rho}}}}}
        \right\}\makebox[\infwidthii]{}}}
\enspace .
\end{displaymath}

\newcommand{\circR}{\circ{\cal R}}

\vskip12pt

\noindent $-/\circR$:
If $\Pi$ is
\begin{displaymath}
\infer[\circR]
{\Seq{B_1,\ldots,B_n,\Gamma}{C}}
  {\left\{\raisebox{-1.5ex}{
      \deduce{\Seq{B_1,\ldots,B_n,\Gamma^i}{C^i}}
                {\Pi^i}}\right\}}
\enspace ,
\end{displaymath}
where $\circR$ is any right rule,
then $\Xi$ reduces to
\settowidth{\infwidthi}
        {$\left\{\raisebox{-2.45ex}{
          \infer[\mc]{\Seq{\Delta_1,\ldots,\Delta_n,\Gamma^i}{C^i}}
                {\deduce{\Seq{\Delta_1}{B_1}}
                        {\Pi_1}
                & \cdots
                & \deduce{\Seq{\Delta_n}{B_n}}
                        {\Pi_n}
                & \deduce{\Seq{B_1,\ldots,B_n,\Gamma^i}{C^i}}
                        {\Pi^i}}}\right\}$}
\settowidth{\infwidthii}{\mc}
\begin{displaymath}
\infer[\circR]
{\Seq{\Delta_1,\ldots,\Delta_n,\Gamma}{C}}
     {\deduce{\makebox[\infwidthi]{}}
            {\left\{\raisebox{-2.45ex}{
            \infer[\mc]{\Seq{\Delta_1,\ldots,\Delta_n,\Gamma^i}{C^i}}
                {\deduce{
                   \Seq{\Delta_1}{B_1}}
                        {\Pi_1}
                & \cdots
                & \deduce{\Seq{\Delta_n}{B_n}}
                        {\Pi_n'}
                & \deduce{\Seq{B_1,\ldots,B_n,\Gamma^i}{C^i}}
                        {\Pi^i}}}\right\}\makebox[\infwidthii]{}}}
\enspace .
\end{displaymath}

\vskip12pt

\noindent{\em \underline{Multicut cases:}}

\vskip12pt

\noindent $\mc/\circL$:
If $\Pi$ ends with a left rule other than $\cL$ acting on $B_1$ and
$\Pi_1$ ends with a multicut and reduces to $\Pi_1'$,
then $\Xi$ reduces to
\begin{displaymath}
\infer[\mc]{\Seq{\Delta_1,\ldots,\Delta_n,\Gamma}{C}}
        {\deduce{\Seq{\Delta_1}{B_1}}
                {\Pi_1'}
        & \deduce{\Seq{\Delta_2}{B_2}}
                {\Pi_2}
        & \cdots
        & \deduce{\Seq{\Delta_n}{B_n}}
                {\Pi_n}
        & \deduce{\Seq{B_1,\ldots,B_n,\Gamma}{C}}
                {\Pi}}
\enspace .
\end{displaymath}

\noindent $-/\mc$:
Suppose $\Pi$ is
\begin{displaymath}
\infer[\mc]{\Seq{B_1,\ldots,B_n,\Gamma^1,\ldots,\Gamma^m,\Gamma'}{C}}
        {\left\{\raisebox{-1.5ex}{\deduce{\Seq{\{B_i\}_{i \in I^j},\Gamma^j}{D^j}}
                {\Pi^j}}\right\}_{j \in \{1..m\}}
        & \raisebox{-2.5ex}{\deduce{\Seq{\{D^j\}_{j \in \{1..m\}},\{B_i\}_{i \in I'},\Gamma'}{C}}
                {\Pi'}}}
\enspace ,
\end{displaymath}
where $I^1,\ldots,I^m,I'$ partition the formulas
$\{B_i\}_{i \in \{1..n\}}$ among the premise derivations
$\Pi_1$, \ldots, $\Pi_m$,$\Pi'$.
For $1 \leq j \leq m$ let $\Xi^j$ be
\begin{displaymath}
\infer[\mc]{\Seq{\{\Delta_i\}_{i \in I^j},\Gamma^j}{D^j}}
        {\left\{\raisebox{-1.5ex}{\deduce{\Seq{\Delta_i}{B_i}}
                        {\Pi_i}}\right\}_{i \in I^j}
        & \raisebox{-2.5ex}{\deduce{\Seq{\{B_i\}_{i \in I^j},\Gamma^j}{D^j}}
                {\Pi^j}}}
\enspace .
\end{displaymath}
Then $\Xi$ reduces to
\begin{displaymath}
\infer[\mc]{\Seq{\Delta_1,\ldots,\Delta_n,\Gamma^1,\ldots\Gamma^m,\Gamma'}{C}}
        {\left\{\raisebox{-1.5ex}{\deduce{\Seq{\ldots}{D^j}}
                        {\Xi^j}}\right\}_{j \in \{1..m\}}
        & \left\{\raisebox{-1.5ex}{\deduce{\Seq{\Delta_i}{B_i}}
                        {\Pi_i}}\right\}_{i \in I'}
        & \raisebox{-2.5ex}{\deduce{\Seq{\ldots}{C}}
                {\Pi'}}}
\enspace .
\end{displaymath}

\vskip12pt

\noindent{\em \underline{Structural case:}}

\vskip12pt

\noindent $-/\cL$:
If $\Pi$ is
\begin{displaymath}
\infer[\cL]{\Seq{B_1,B_2,\ldots,B_n,\Gamma}{C}}
        {\deduce{\Seq{B_1,B_1,B_2,\ldots,B_n,\Gamma}{C}}
                {\Pi'}}
\enspace ,
\end{displaymath}
then $\Xi$ reduces to
\settowidth{\infwidthi}
        {$\Seq{\Delta_1,\Delta_1,\Delta_2,\ldots,\Delta_n,\Delta_n,\Gamma}{C}$}
\begin{displaymath}
\infer{\Seq{\Delta_1,\Delta_2,\ldots,\Delta_n,\Gamma}{C}}
        {\infer[\cL]{\makebox[\infwidthi]{}}
                {\infer[\mc]{\Seq{\Delta_1,\Delta_1,\Delta_2,\ldots,\Delta_n,\Delta_n,\Gamma}{C}}
                        {\raisebox{-2.5ex}{\deduce{\Seq{\Delta_1}{B_1}}
                                {\Pi_1}}
                        & \left\{\raisebox{-1.5ex}{\deduce{\Seq{\Delta_i}{B_i}}
                                        {\Pi_i}}\right\}_{i \in \{1..n\}}
                        & \raisebox{-2.5ex}{\deduce{\Seq{B_1,B_1,B_2,\ldots,B_n,\Gamma}{C}}
                                {\Pi'}}}}}
\enspace .
\end{displaymath}

\vskip12pt

\noindent{\em \underline{Axiom cases:}}

\vskip12pt

\noindent $id_\pi/\circL$:
Suppose $\Pi$ ends with either $\natL$ or $\eqL$ on $B_1$ and 
$\Pi_1$ ends with the $id_\pi$ rule:
$$
\infer[id_\pi]
{\Seq{\Delta_1',B}{B_1}}
{\pi_1.B = \pi_2.B_1}
$$
Then it is the case that $B = \pi_1^{-1}.\pi_2.B_1.$
Apply the construction in Definition~\ref{def:perm}
to $\Pi$ to get a derivation $\Pi'$ of
$\Seq{B, B_2,\ldots,B_n,\Gamma}{C}.$
The derivation $\Xi$ reduces to
\begin{displaymath}
\infer[\mc]{\Seq{B, \Delta_1', \Delta_2,\ldots,\Delta_n,\Gamma}{C}}
        {\deduce{\Seq{\Delta_2}{B_2}}
                {\Pi_2}
        & \cdots
        & \deduce{\Seq{\Delta_n}{B_n}}
                {\Pi_n}
        & \deduce{\Seq{B,\Delta_1', B_2,\ldots,B_n,\Gamma}{C}}
                {w(\Delta_1',\Pi')}}          
\enspace .
\end{displaymath}

\vskip12pt

\noindent $-/id_\pi$:
If $\Pi$ ends with the $id_\pi$ rule with a matching
formula in $\Gamma$, i.e., there exists $C' \in \Gamma$
such that $\pi.C' = \pi'.C$ for some permutations
$\pi$ and $\pi'$, then 
then $\Xi$ reduces to
$$
\infer[id_\pi]
{\Seq{\Delta_1,\ldots,\Delta_n,\Gamma}{C}}{}
$$
If $\Pi$ ends with the $id_\pi$ rule but $C$ does not
match any formula in $\Gamma$, then $C$ must match one
of the cut formulas, say $B_1$, i.e.,
there exists permutations $\pi_1$ and $\pi_2$ such
that $\pi_1.B_1 = \pi_2.C$. That is, $C = \pi_2^{-1}.\pi_1.B_1.$
In this case, we first apply the permutation 
$\pi_2^{-1}.\pi_1$ to $\Pi_1$ according to the construction
in Definition~\ref{def:perm} to get a derivation
$\Pi_1'$ of $\Seq{\Delta_1}{\pi_2^{-1}.\pi_1.B_1}$. 
$\Xi$ then reduces to  
$w(\Delta_2\cup \ldots \cup \Delta_n \cup \Gamma, \Pi_1').$
\qed

An inspection of the rules of the logic and this definition will
reveal that every derivation ending with a multicut has a
reduct.
Because we use a multiset as the left side of the sequent, there may
be ambiguity as to whether a formula occurring on the left side of the
rightmost premise to a multicut rule is in fact a cut formula, and if
so, which of the left premises corresponds to it.
As a result, several of the reduction rules may apply, and so a
derivation may have multiple reducts.

\newcommand{\Ascr}{{\cal A}}
\newcommand{\Bscr}{{\cal B}}
\newcommand{\Cscr}{{\cal C}}
\newcommand{\Dscr}{{\cal D}}
\newcommand{\Escr}{{\cal E}}
\newcommand{\Fscr}{{\cal F}}
\newcommand{\Gscr}{{\cal G}}
\newcommand{\Hscr}{{\cal H}}
\newcommand{\ul}[1]{\underline{#1}}
\newcommand{\Tscr}{{\cal T}}

\section{Normalizability and reducibility}
\label{sec:norm}

We now define two properties of derivations:  normalizability and
reducibility.
Each of these properties implies that the derivation can be reduced to
a cut-free derivation of the same end-sequent.
In the following, substitutions mean $\Sigma$-substitutions for
some signature $\Sigma.$
The definitions are similar to those by McDowell and 
Miller~\cite{mcdowell00tcs}. 
However, since the cut reduction in our case involves 
several transformations of derivations, other than substitutions
and weakening, we need to build this transformations
into the definitions of normalizability and reducibility.

\begin{definition}
A {\em height-preserving} (HP) transformation ${\cal T}$
is a finite sequence of transformations ${\cal F}_1, \ldots , {\cal F}_n$
where each ${\cal F}_i$ is one of the transformations
described in Definition~\ref{def:weak}, Definition~\ref{def:subst},
Definition~\ref{def:perm} and Definition~\ref{def:res}.
The number $n$ is the {\em order} of ${\cal T}$.
The application of ${\cal T}$ to $\Pi$ is defined as follows:
$$
\begin{array}{ll}
{\cal T}_0(\Pi) = \Pi\\
{\cal T}_{i+1}(\Pi) = {\cal F}_{i+1}({\cal T}_i(\Pi))\\
{\cal T}(\Pi) = {\cal T}_n(\Pi)
\end{array}
$$
\end{definition}
Note that a height-preserving transformation may not be defined for all derivations, and that
it may be the identity transformation (i.e., it does nothing).
Height-preserving transformations are ranged over by $\Tscr, \Fscr, \Gscr$
and $\Hscr.$

\begin{lemma}
\label{lm:height}
Let ${\cal T}$ be a height-preserving transformation.
For any derivation $\Pi$, if ${\cal T}(\Pi)$ is defined,
then $ht({\cal T}(\Pi)) \leq ht(\Pi).$
\end{lemma}

\begin{definition}
\label{def:norm}
We define the set of {\em normalizable} derivations to be the smallest
set that satisfies the following conditions:
\begin{enumerate}
\item
If a derivation $\Pi$ ends with a multicut, then it is normalizable
if for every height-preserving transformation ${\cal T}$ such that ${\cal T}(\Pi)$ is defined,
there is a normalizable reduct of ${\cal T}(\Pi)$.

\item
If a derivation ends with any rule other than a multicut,
then it is normalizable if the premise derivations are normalizable.
\end{enumerate}
These clauses assert that a given derivation is normalizable
provided certain (perhaps infinitely many) other derivations are
normalizable.
If we call these other derivations the predecessors of the given
derivation, then a derivation is normalizable if and only if the tree
of the derivation and its successive predecessors is well-founded.
In this case, the well-founded tree is called the {\em normalization} of
the derivation.
\end{definition}
The set of normalizable derivations is not empty; the cut-free
proofs, for instance, are normalizable.

Since a normalization is well-founded, it has an associated induction
principle:  for any property $P$ of derivations, if for every
derivation $\Pi$ in the normalization, $P$ holds for every predecessor
of $\Pi$ implies that $P$ holds for $\Pi$, then $P$ holds for every
derivation in the normalization.

\begin{lemma}
\label{lm:norm-cut-free}
If there is a normalizable derivation of a sequent, then there is a
cut-free derivation of the sequent.
\end{lemma}
\begin{proof} 
Let $\Pi$ be a normalizable derivation of the sequent
$\Seq{\Gamma}{B}$.
We show by induction on the normalization of $\Pi$ that there is a
cut-free derivation of $\Seq{\Gamma}{B}$.
\begin{enumerate}
\item
If $\Pi$ ends with a multicut, then any of its reducts is one of its
predecessors  and so is normalizable.
One of its reduct, via the empty transformation, is also 
a derivation of $\Seq{\Gamma}{B}$, so by the
induction hypothesis this sequent has a cut-free derivation.

\item
Suppose $\Pi$ ends with a rule other than multicut.
Since we are given that $\Pi$ is normalizable, by definition the
premise derivations are normalizable.
These premise derivations are the predecessors of $\Pi$, so by the
induction hypothesis there are cut-free derivations of the premises.
Thus there is a cut-free derivation of $\Seq{\Gamma}{B}$.
\end{enumerate}
\qed
\end{proof}

The next four lemmas are also proved by induction on the
normalization of derivations.

\begin{lemma}
\label{lm:norm subst}
If $\Pi$ is a normalizable derivation, then for any substitution
$\theta$ such that $\Pi\theta$ is defined,  
$\Pi\theta$ is normalizable.
\end{lemma}

\begin{lemma}
\label{lm:norm weak}
If $\Pi$ is normalizable, then for any multiset of formulas $\Delta$,
if $w(\Delta,\Pi)$ is defined, then $w(\Delta,\Pi)$ is normalizable.
\end{lemma}

\begin{lemma}
\label{lm:norm perm}
If $\Pi$ is normalizable, then for any permutations $\vec \pi$
such that $\langle \vec \pi \rangle.\Pi$ is defined,
$\langle \vec \pi \rangle.\Pi$ is normalizable.
\end{lemma}

\begin{lemma}
\label{lm:norm res}
If $\Pi$ is normalizable, then for any nominal constants $\vec a$
such that $r(x,\vec a,\Pi)$ is defined, 
$r(x,\vec a,\Pi)$ is normalizable.
\end{lemma}

\begin{lemma}
\label{lm:norm hpt}
If $\Pi$ is normalizable, then for any height-preserving transformation
${\cal T}$ such that ${\cal T}(\Pi)$ is defined,
${\cal T}(\Pi)$ is normalizable.
\end{lemma}

\begin{definition}
\label{def:level drv}
The level of a sequent $\Seq \Gamma C$ is the level of $C$.
The level of a derivation $\Pi$ is the level of its root sequent.
\end{definition}

The definition of reducibility for derivations 
is done by induction on the level of derivations:
in defining the reducibility of level-$i$ derivations,
we assume that the reducibility of derivations
of level $j$, for all $j < i$ is already defined.
In the following definition,
when we apply a transformation ${\cal T}$ to
a derivation $\Pi$ of $\Seq{B_1,\dots,B_n}{B_0},$
we use the notation ${\cal T}(B_i)$ to denote 
the formula in the root sequent of ${\cal T}(\Pi)$ 
that results from applying the transformation to $B_i$.

\begin{definition}
\label{def:reducibility}
{\em Reducibility.}
For any $i$, we define the set of {\em reducible}
$i$-level derivations to be the smallest
set of $i$-level derivations that satisfies
the following conditions:
\begin{enumerate}
\item If a derivation $\Pi$ ends with a multicut then it is reducible
if for every height-preserving transformation ${\cal T}$ such
that ${\cal T}(\Pi)$ is defined,
there is a reducible reduct of ${\cal T}(\Pi).$
\item Suppose the derivation ends with the implication right rule
$$
\infer[\oimpR]
{\Seq{\Gamma}{B\oimp C}}
{\deduce{\Seq{B,\Gamma}{C}}{\Pi}}
$$
Then the derivation is reducible if $\Pi$ is reducible and
for every height-preserving transformation ${\cal T}$ such that
${\cal T}(\Pi)$ is defined, multiset of formulas $\Delta$ and
reducible derivation $\Pi'$ of $\Seq{\Delta}{B'}$, where $B' = {\cal T}(B)$,
the derivation
$$
\infer[mc]
{\Seq{\Delta, \Gamma'}{C'}}
{\deduce{\Seq \Delta {B'}}{\Pi'}
& \deduce{\Seq{B',\Gamma'}{C'}}{{\cal T}(\Pi)}
}
$$
is reducible.
\item If the derivation ends with the implication left rule or the $nat$ rule,
then it is reducible if the right premise derivation is reducible and the
other premise derivations are normalizable.
\item If the derivation ends with any other rule, then it is reducible if
the premise derivations are reducible.
\end{enumerate}
These clauses assert that a given derivation is reducible 
provided certain other derivations 
are reducible. If we call these other derivations the 
predecessors of the given derivation, then a derivation 
is reducible only if the tree of the derivation and its 
successive predecessors is well founded. 
In this case, the well founded tree is called the {\em reduction}
of the derivation.
\end{definition}

\begin{lemma}
\label{lm:reducible implies norm}
If a derivation is reducible, then it is normalizable.
\end{lemma}
\begin{proof} By induction on the reduction of the derivation. 
\qed
\end{proof}

\begin{lemma}
\label{lm:reducible hpt}
If a derivation $\Pi$ is reducible, then for
any height-preserving ${\cal T}$ such
that ${\cal T}(\Pi)$ is defined, ${\cal T}(\Pi)$
is reducible.
\end{lemma}
\begin{proof}
By induction on the reduction of $\Pi$ and
Lemma~\ref{lm:norm hpt}.
\end{proof}

\section{Cut elimination}
\label{sec:cut-elim}

In the following, when we mention ${\cal T}(\Pi)$ we assume
implicitly that it is defined. 
We shall also use the notation $\underline{B}_{\cal T}$
to denote ${\cal T}(B)$, that is the application of
the transformation to the formula $B.$ Similarly, the multiset ${\cal T}(\Delta)$
will be written $\underline{\Delta}_{\cal T}.$
We drop the subscript ${\cal T}$ if it is clear from context
which transformation we refer to.

\begin{lemma}
\label{lm:reducibility}
For any derivation $\Pi$ of $\NSeq{\Sigma}{B_1,\ldots,B_n,\Gamma}{C}$
and reducible derivations $\Pi_1, \ldots, \Pi_n$
of $\NSeq{\Sigma}{\Delta_1}{C_1}, \ldots, \NSeq{\Sigma}{\Delta_n}{C_n},$
where $n \geq 0$, 
and for any transformations $\Tscr_1, \ldots, \Tscr_n, \Tscr$
such that $\Tscr_i(\Pi_i)$ is defined and 
$\Tscr_i(C_i) = \Tscr(B_i)$, 
the derivation $\Xi$
$$
\infer[\mc]
{\NSeq{\Sigma'}{\ul{\Delta_1}_{\Tscr_1},\ldots,
\ul{\Delta_n}_{\Tscr_n},\ul{\Gamma}_{\Tscr}}{\ul{C}_{\Tscr}}}
{
\deduce{\NSeq{\Sigma'}{\ul{\Delta_1}_{\Tscr_1}}{\ul{B_1}_{\Tscr}}}{\Tscr_1(\Pi_1)} &
\ldots &
\deduce{\NSeq{\Sigma'}{\ul{\Delta_n}_{\Tscr_n}}{\ul{B_n}_{\Tscr}}}{\Tscr_n(\Pi_n)} &
\deduce{\NSeq{\Sigma'}{\ul{B_1}_{\Tscr},\ldots,\ul{B_n}_{\Tscr}, \ul{\Gamma}_{\Tscr}}
{\ul{C}_{\Tscr}}}{\Tscr(\Pi)}
}
$$
is reducible.
\end{lemma}
\begin{proof}
The proof is by induction on $ht(\Pi)$ with subordinate induction
on $n$ and on the reductions of $\Pi_1,\ldots,\Pi_n.$
Since the proof does not depend on the order of the inductions
on reductions, when we need to distinguish of one the $\Pi_i$'s
we shall refer to it as $\Pi_1$ without loss of generality.

We need to show that for every $\Tscr'$, the derivation
every reduct of $\Tscr'(\Xi)$ is reducible.
If $n=0$ then $\Tscr'(\Xi)$ reduces to $\Tscr'(\Tscr(\Pi)).$
Since reducibility is preserved by height-preserving
transformation, it suffices to consider the case where
$\Tscr$ and $\Tscr'$ are the identity transformation, that is,
we need only to show that $\Pi$ is reducible.
This is proved by case analysis on the last rule of
$\Pi.$
For each case, the results follow from the outer induction 
hypothesis and Definition~\ref{def:reducibility}. 
The case with $\oimpR$ requires that height-preserving
transformations do not increase the height of the derivations
(see Lemma~\ref{lm:height}).
In the cases for $\oimpL$ and $\natL$ we need the
additional information that reducibility implies
normalizability (see Lemma~\ref{lm:reducible implies norm}).

For $n > 0$, we analyze all possible reductions that apply to $\Tscr'(\Xi)$ 
and show that every reduct of ${\cal T}'(\Xi)$ is reducible. 
We suppose that $\Tscr'(\Xi)$ is of the following form:
$$
\infer[mc]
{\Seq{\ul{\Delta_1}_{\Fscr_1}, \ldots, \ul{\Delta_n}_{\Fscr_n},
\ul{\Gamma}_{\Fscr}}{\ul{C}_{\Fscr}}}
{
\deduce{\Seq{\ul{\Delta_1}_{\Fscr_1}}{\ul{C_1}_{\Fscr_1}}}{\Fscr_1(\Pi_1)}
&
\ldots
&
\deduce{\Seq{\ul{\Delta_n}_{\Fscr_n}}{\ul{C_n}_{\Fscr_n}}}{\Fscr_n(\Pi_n)}
&
\deduce{\Seq{\ul{B_1}_{\Fscr},\ul{B_n}_{\Fscr}, \ul{\Gamma}_{\Fscr}}
{\ul{C}_{\Fscr}}}{\Fscr(\Pi)}
}
$$
where $\ul{B_i}_{\Fscr} = \ul{C_i}_{\Fscr_i}.$
In several cases below, we often omit the subscripts $\Fscr$ or $\Fscr_i$
when it is clear from context which transformations we refer to.
We also often switch between $\ul{B_i}_{\Fscr}$ and $\ul{C_i}_{\Fscr_i}$
to make the inference figures more readable.
 
Most cases follow immediately from the inductive
hypothesis and Definition~\ref{def:reducibility} and 
Lemma~\ref{lm:reducible implies norm}, Lemma~\ref{lm:reducible hpt} and 
Lemma~\ref{lm:height}. We show here the interesting cases. 

\vskip12pt
\noindent $\oimpR/\oimpL$: Suppose $\Pi_1$ and $\Pi$ are
$$
\infer[\oimpR]
{\Seq{\Delta_1}{B_1'\oimp B_1''}}
{
\deduce{\Seq{\Delta_1,B_1'}{B_1''}}{\Pi_1'}
}
\qquad \qquad
\infer[\oimpL]
{\Seq{B_1'\oimp B_1'',B_2,\dots, B_n,\Gamma}{C}}
{
\deduce{\Seq{B_2,\dots,\Gamma}{B_1'}}{\Pi'}
&
\deduce{\Seq{B_1'',B_2,\dots,\Gamma}{C}}{\Pi''}
}
\enspace .
$$
Let $\Xi_1$ be the derivation
$$
\infer[\mc]
{\Seq{\underline{\Delta_2},\dots,\underline{\Delta_n},\underline{\Gamma}}{\underline{B_1'}}}
{
\deduce{\Seq{\underline {\Delta_2}}{\underline{B_2}}}{\Fscr_2(\Pi_2)}
&
\ldots
&
\deduce{\Seq{\underline{\Delta_n}}{\underline{B_n}}}{\Fscr_n(\Pi_n)}
&
\deduce{\Seq{\underline{B_2},\dots,\underline{B_n}, \underline \Gamma}{\underline{B_1'}}}
             {\Fscr_n(\Pi')}
}
$$
Then $\Xi_1$ is reducible by induction hypothesis 
since $\Fscr$ and $\Fscr_i$ preserve
reducibility (Lemma~\ref{lm:reducible hpt}) and do not increase
the height of derivations (Lemma~\ref{lm:height}). Since we are given that
$\Pi_1$ is reducible, by Definition~\ref{def:reducibility}, the derivation
$\Xi_2$
$$
\infer[\mc]
{\Seq{\ul{\Delta_1}, \ldots, \ul{\Delta_n}, \ul \Gamma}{\ul {B_1''}}}
{
\deduce{\Seq{\ul{\Delta_2}, \ldots, \ul{\Delta_n}, \ul\Gamma}
  {\ul {B_1'}}}{\Xi_1}
&
\deduce{\Seq{\ul{B_1'}, \ul {\Delta_1}}{\ul {B_1''}}}{\Fscr_1(\Pi_1')}
}
$$
is reducible as well.
Therefore, the reduct of $\Tscr'(\Xi)$
\settowidth{\infwidthi}
        {$\Seq{\underline{\Delta_1},\ldots,\underline{\Delta_n},
\underline{\Gamma},\underline{\Delta_2},\ldots,\underline{\Delta_n},\underline \Gamma}{\underline C}$}
\begin{displaymath}
\infer{\Seq{\underline{\Delta_1},\ldots,\underline{\Delta_n},\underline{\Gamma}}{\underline C}}
        {\infer[\cL]{\makebox[\infwidthi]{}}
                {\infer[\mc]{\Seq{\ul{\Delta_1},\ldots,\ul{\Delta_n}, \ul\Gamma,
                                        \ul{\Delta_2},\ldots,\ul{\Delta_n},\ul{\Gamma}}{\ul C}}
                        {\raisebox{-2.5ex}{\deduce{\Seq{\ldots}{\ul{B_1''}}}
                                {\Xi_2}}
                        & \left\{\raisebox{-1.5ex}{\deduce{\Seq{\ul{\Delta_i}}{\ul{B_i}}}
                                {\Fscr_i(\Pi_i)}}\right\}_{i \in \{2..n\}}
                        & \raisebox{-2.5ex}{
     \deduce{\Seq{\ul{B_1''},\{\ul{B_i}\}_{i \in \{2..n\}},\ul \Gamma}{\ul C}}
                                {\Fscr(\Pi'')}}}}}
\enspace .
\end{displaymath}
is reducible by the outer induction hypothesis and Definition~\ref{def:reducibility}.

\vskip12pt
\noindent $\forallR/\forallL:$
Suppose $\Pi_1$ and $\Pi$ are
$$
\infer[\forallR]
{\NSeq{\Sigma}{\Delta_1}{\forall x.B}}
{\deduce{\NSeq {\Sigma, h} {\Delta_1} {B[h\,\vec c/x]}}{\Pi_1'}}
\qquad
\infer[\forallL]
{\NSeq{\Sigma}{\forall x.B, B_2, \ldots, B_n,\Gamma}{C}}
{
\deduce{\NSeq{\Sigma}{B[t/x], B_2,\ldots, B_n,\Gamma}{C}}{\Pi'}
}
$$
Applying the transformation $\Fscr_1$ to $\Pi_1$ (and similarly,
$\Fscr$ to $\Pi$) might require several transformation be done
on the premise of the derivation, e.g., to avoid clashes of 
nominal constants, etc., so let us suppose that $\Fscr_1(\Pi_1)$
and $\Fscr(\Pi)$ are of the following shapes:
$$
\infer[\forallR]
{\NSeq{\Sigma'}{\ul{\Delta_1}}{\forall x.D}}
{\deduce{\NSeq {\Sigma', h} {\Delta_1} {D[h'\,\vec d/x]}}{\Gscr_1(\Pi_1')}}
\qquad
\infer[\forallL]
{\NSeq{\Sigma'}{\forall x.D, \ul{B_2}, \ldots, \ul{B_n}, \ul{\Gamma}}{\ul{C}}}
{
\deduce{\NSeq{\Sigma'}{D[s/x], \ul{B_2},\ldots, \ul{B_n}, \ul{\Gamma}}{\ul C}}
{\Gscr(\Pi')}
}
$$
where $\forall x.D = \ul{\forall x.B}$ and $D[s/x] = \ul{B[t/x]}.$
If the support of $D[s/x]$ is larger than $\{ \vec d \}$,
then the reduction rule for $\forallR/\forallL$ requires further
transformations be applied to $\Gscr_1(\Pi_1')$, i.e.,
as is described in Lemma~\ref{lm:supp1}.
So let us suppose that this transformation is applied, resulting
in a derivation 
$$
\deduce{\NSeq{\Sigma',f}{\ul{\Delta_1}}{D[f\vec e/x]}}{\Gscr_1'(\Pi_1')}
\enspace .
$$
Then $\Tscr'(\Xi)$ reduces to
$$
\infer[mc]
{\NSeq{\Sigma'}{\ul{\Delta_1},\ldots,\ul{\Delta_n}, \ul \Gamma}{\ul C}}
{\deduce{\NSeq{\Sigma'}{\ul {\Delta_1}}{D[s/x]}}
 {\Gscr_1'(\Pi_1')[\lambda \vec e.s/f]}
&
\deduce{\Seq{\ul{\Delta_2}}{\ul{B_2}}}{\Fscr_2(\Pi_2)}
&
\ldots
&
\deduce{\Seq{\ul{\Delta_2}}{\ul{B_2}}}{\Fscr_n(\Pi_n)}
&
\deduce{\NSeq{\Sigma'}{D[s/x], \ldots, \ul{\Gamma}}{\ul C}}{\Gscr(\Pi')}
}
$$
which is reducible by the outer induction hypothesis.

\vskip12pt

\noindent $\natR/ \natL:$
Suppose $\Pi_1$ and $\Pi$ are 
$$
\infer[\natR]
{\Seq{\Delta_1}{nat\,M}}
{\deduce{\Seq{\Delta_1}{nat\,M}}{\Pi_1'}}
\qquad
\infer[\natL]
{\Seq{nat\,(s\,I), B_2, \ldots, B_n,\Gamma}{C}}
{\deduce{\Seq{}{D\,z}}{\Pi'}
& \deduce{\Seq{D\,j}{D\,(s\,j)}}{\Pi''}
& \deduce{\Seq{D\,(s\,M),B_2,\ldots,B_n,\Gamma}{C}}{\Pi'''}
}
$$
then $\Fscr_1(\Pi_1)$ and $\Fscr(\Pi)$ are
$$
\infer[\natR]
{\Seq{\ul{\Delta_1}}{nat\,I}}
{\deduce{\Seq{\ul{\Delta_1}}{nat\,I}}{\Fscr_1(\Pi_1')}}
\qquad
\infer[\natL]
{\Seq{nat\,(s\,I), \ul{B_2}, \ldots, \ul{B_n},\ul{\Gamma}}{\ul{C}}}
{\deduce{\Seq{}{D\,z}}{\Pi'}
& \deduce{\Seq{D\,j}{D\,(s\,j)}}{\Pi''}
& \deduce{\Seq{D\,(s\,I),\ul{B_2},\ldots,\ul{B_n},\ul \Gamma}{\ul C}}{\Fscr(\Pi''')}
}
$$
Note that the derivations $\Pi'$ and $\Pi''$ are not affected
by the transformation $\Fscr$ since $D$ is a closed term
with no occurrences of nominal constants and $j$ in
$\Pi''$ is a new eigenvariable.
Let $\Xi_1$ be the derivation
$$
\infer[mc]
{\Seq{\ul{\Delta_1}}{D\,I}}
{
\deduce{\Seq{\ul {\Delta_1}}{nat\,I}}{\Fscr_1(\Pi_1')}
& 
\infer[\natL]
{\Seq{nat\,I}{D\,I}}
{
\deduce{\Seq{}{D\,z}}{\Pi'} &
\deduce{\Seq{D\,j}{D\,(s\,j)}}{\Pi''}
&
\infer[id_\pi]
{\Seq{D\,I}{D\,I}}
{}
}
}
\enspace .
$$
Since the height of the right premise is no larger than $ht(\Pi)$,
and $\Pi_1'$ is a predecessor of $\Pi_1$, 
$\Xi_1$ is reducible by induction on the reduction of $\Pi_1.$
Let $\{\vec c\}$ be the support of $I.$
We construct the derivation $\Pi^\bullet$ of
$\NSeq{h}{D\,(h\,\vec c)}{D\,(s\,(h\,\vec c))}$
from $\Pi''$ using the procedures described in Definition~\ref{def:subst}
and Definition~\ref{def:res}. 
Let $\Xi_2$ be
$$
\infer[\mc .]
{\Seq{\ul{\Delta_1}}{D\,(s\, I)}}
{
\deduce{\Seq{\ul{\Delta_1}}{D\,I}}{\Xi_1}
&
\deduce{\Seq{D\,I}{D\,(s\,I)}}{\Pi^{\bullet}[\lambda \vec c.I/h]}
}
$$
Since $ht(\Pi^{\bullet} [ \lambda \vec c.I/h ]) \leq ht( \Pi'')$,
by the outer induction hypothesis, $\Xi_2$ is also reducible.
Therefore the reduct of $\Tscr'(\Xi)$
$$
\infer[mc]
{\Seq{\ul{\Delta_1}, \ldots, \ul{\Delta_2},\ul \Gamma}{\ul C}}
{
\deduce{\Seq{\ul {\Delta_1}}{D\,(s\,I)}}{\Xi_2}
&
\deduce{\Seq{\ul {\Delta_2}}{\ul{B_2}}}{\Fscr_2(\Pi_2)}
&
\ldots
&
\deduce{\Seq{\ul {\Delta_n}}{\ul{B_n} } }{\Fscr_n(\Pi_n)}
&
\deduce{\Seq{D\,(s\,I), \ul{B_2}, \ldots, \ul \Gamma}{\ul C}}{\Fscr(\Pi''')}
}
$$
is reducible by the outer induction hypothesis.

\vskip12pt

\noindent $\eqL/\circL:$
Suppose $\Pi_1$ is 
$$
\infer[\eqL]
{\Seq{s = t, \Delta_1}{B_1}}
{
\left\{
\raisebox{-1.5ex}{\deduce{\Seq{\Delta_1\theta}{B_1\theta}}{\Pi^\theta}}
\right\}_\theta
}
$$
then $\Fscr_1(\Pi_1)$ is 
$$
\qquad
\infer[\eqL]
{\Seq{\ul s = \ul t, \ul{\Delta_1}}{\ul{B_1}}}
{
\left\{
\raisebox{-1.5ex}{\deduce{\Seq{\ul{\Delta_1}\theta}{\ul{B_1}\theta}}{\Pi^{\bullet\rho}}}
\right\}_\rho
}
$$
where each $\Pi^{\bullet\rho}$ is obtained from some $\Pi^\theta$ by
the transformations described in Definition~\ref{def:weak},
Definition~\ref{def:subst}, Definition~\ref{def:perm}
and Definition~\ref{def:res}. 
We denote with $f(\rho)$ the substitution $\theta$ such that 
$\Pi^{\bullet\rho}$ is constructed out of $\Pi^\theta.$
Thus we can write each $\Pi^{\bullet\rho}$
as the derivation $\Fscr_\rho(\Pi^{f(\rho)})$ for some
transformation $\Fscr_\rho.$
The reduct of $\Tscr'(\Xi)$ 
$$
\infer[\eqL]
{
\Seq{\qquad \qquad \qquad \quad
\ul s= \ul t,\ul{\Delta_1'},\ul{\Delta_2},\ldots,\ul{\Delta_n},\ul \Gamma}{\ul C
\qquad \qquad \qquad \qquad \qquad \qquad \qquad\qquad}
}
{
\left\{
\raisebox{-1.5ex}
{
\infer[mc]
{\Seq{\ul{\Delta_1'}\rho, \ul{\Delta_2}\rho,\ldots, \ul{\Delta_n}\rho,
\ul \Gamma\rho}{\ul C \rho}}
{
\deduce{\Seq{\ul{\Delta_1'}\rho}{\ul{B_1}\rho}}{\Fscr_\rho(\Pi^{f(\rho)})}
&
\deduce{\Seq{\ul{\Delta_2}\rho}{\ul{B_2}\rho}}{\Fscr_2(\Pi_2)\rho}
&
\ldots
&
\deduce{\Seq{\ul{\Delta_n}\rho}{\ul{B_n}\rho}}{\Fscr_n(\Pi_n)\rho}
&
\deduce{\Seq{\ul{B_1}\rho,\ldots,\ul{B_n}\rho,\ul{\Gamma}\rho}{\ul C\rho}}{\Fscr(\Pi)\rho}
}
}
\right\}_\rho
}
$$
Each premise derivation of the above derivation is reducible by
the induction hypothesis on the reduction of $\Pi_1$, since each
$\Pi^{f(\rho)}$ is a predecessor of $\Pi_1.$
The reduct of $\Tscr'(\Xi)$ is therefore reducible by 
Definition~\ref{def:reducibility}.

\vskip12pt

\noindent $-/\oimpR:$
Suppose $\Pi$ is 
$$
\infer[\oimpR]
{\Seq{B_1, \ldots, B_n, \Gamma}{ {C_1 \oimp C_2}}}
{
\deduce{\Seq{B_1, \ldots, B_n, \Gamma, C_1 }{C_2}}{\Fscr(\Pi')}
}
$$
then $\Fscr_1(\Pi)$
$$
\infer[\oimpR]
{\Seq{\ul{B_1}, \ldots, \ul{B_n}, \ul \Gamma}{\ul {C_1 \oimp C_2}}}
{
\deduce{\Seq{\ul{B_1}, \ldots, \ul{B_n}, \ul \Gamma, \ul{C_1} }{\ul{C_2}}}{\Fscr(\Pi')}
}
$$
Let $\Xi_1$ be
$$
\infer[]
{\Seq{\ul{\Delta_1}, \ldots, \ul{\Delta_n}, \ul{C_1}}{\ul{C_2}}}
{
\deduce{\Seq{\ul{\Delta_1}}{\ul{B_1}}}{\Fscr_1(\Pi_1)}
&
\ldots
&
\deduce{\Seq{\ul{\Delta_n}}{\ul{B_n}}}{\Fscr_n(\Pi_n)}
&
\deduce{\Seq{\ul{B_1},\ldots,\ul{B_1},\ul{\Gamma}, \ul{C_1}}{\ul{C_2}}}{\Fscr(\Pi')}
}
$$
which is reducible by the outer induction hypothesis.
Let $\Xi_2$ be the derivation
$$
\infer[\oimpR]
{\Seq{\ul{\Delta_1}, \ldots, \ul{\Delta_n}, \ul{\Gamma}}
{\ul{C_1\oimp C_2}}}
{
\deduce{\Seq{\ul{\Delta_1}, \ldots, \ul{\Delta_n}, \ul \Gamma, \ul {C_1}}{\ul{C_2}}}{\Xi_1}
}
\enspace ,
$$
which is the reduct of $\Tscr'(\Xi).$
To show that $\Xi_2$ is reducible, we need to show
that for any $\Tscr''$, and for any derivation $\Pi''$ of $\Seq{\Delta}{D},$
where $D = \Tscr''(\ul{C_1})$, the derivation $\Xi_3$
$$
\infer[mc]
{\Seq{\Delta, \ul{\Delta_1}_{\Gscr_1}, \ldots, \ul{\Delta_n}_{\Gscr_n}, \ul{\Gamma}_{\Gscr}}
{\ul{C_2}_{\Gscr}}}
{
\deduce{\Seq{\Delta}{D}}{\Pi''}
&
\deduce{\Seq{D, \ul{\Delta_1}_{\Gscr_1}, \ldots, \ul{\Delta_n}_{\Gscr_n},
      \ul{\Gamma}_{\Gscr}}{\ul{C_2}_{\Gscr}} }{\Tscr''(\Xi_2)}
}
$$
is reducible. Here the transformations $\Gscr_i$ and $\Gscr$ are
transformations associated with the premise derivations 
in $\Tscr''(\Xi_2).$ $\Xi_3$ is reducible if for any transformation
$\Hscr$, every reduct of the derivation $\Hscr(\Xi_3)$ is reducible.
The reduct of $\Hscr(\Xi_3)$ in this case is:
$$
\infer[\mc]
{\Seq{\ul{\Delta}, \ul{\Delta_1}, \ldots, \ul{\Delta_n}, \ul{\Gamma}}{\ul{C_2}}}
{
\deduce{\Seq{\ul{\Delta}}{\ul D}}{\Hscr'(\Pi'')}
&
\deduce{\Seq{\ul {\Delta_1}}{\ul {B_1}}}{\Hscr_1(\Pi_1)}
&
\ldots
&
\deduce{\Seq{\ul {\Delta_n}}{\ul{B_n}}}{\Hscr_n(\Pi_n)}
&
\deduce{\Seq{\ul D, \ul{B_1}, \ldots, \ul{B_n}, \ul{\Gamma}}{\ul{C_2}}}{\Hscr''(\Pi')}
}
$$
where $\Hscr_1,\ldots,\Hscr_n$ and $\Hscr''$ are transformations applied
to the premises of $\Hscr(\Tscr''(\Xi_2))$ and
$\Hscr'$ is the transformation applied to the left premise of
$\Hscr(\Xi_3).$ This derivation is reducible by the outer induction hypothesis.
\qed
\end{proof}

\begin{corollary}
\label{cor:reducibility}
Every derivation is reducible.
\end{corollary}
\begin{proof}
This result follows immediately from Lemma~\ref{lm:reducibility}
with $n = 0.$ 
\qed
\end{proof}

\begin{theorem}
\label{thm:cut elim}
The cut rule is admissible in $\LGN$.
\end{theorem}
\begin{proof}
Follows immediately from Corollary~\ref{cor:reducibility},
Lemma~\ref{lm:reducible implies norm} and Lemma~\ref{lm:norm-cut-free}.\qed
\end{proof}

\begin{corollary}
The logic $\LGN$ is consistent, i.e., it is not the case
that both $A$ and $A \oimp \bot$
are provable.
\end{corollary}

\section{Correspondence between $\LG$ and $\FOLNb$}
\label{sec:corr}

\newcommand{\Judg}[2]{#1 \triangleright #2}

\begin{figure}
{\small
$$
\infer[id]
      {\NSeq{\Sigma}{\Judg{\sigma}{B},\Gamma}{\Judg{\sigma}{B}}}{}
\qquad 
\infer[cut]
        {\NSeq{\Sigma}{\Delta,\Gamma}{\Cscr}}
        {\NSeq{\Sigma}{\Delta}{\Bscr} \qquad
         \NSeq{\Sigma}{\Bscr,\Gamma}{\Cscr}}
$$
$$
\infer[\landL]
      {\NSeq
        {\Sigma}
        {\Judg{\sigma}{B \land C},\Gamma}
        {\Dscr}
      }
      {\NSeq
        {\Sigma}
        {\Judg{\sigma}{B}, \Judg{\sigma}{C}, \Gamma}
        {\Dscr}
      }
\qquad
\infer[\landR]
      {\NSeq{\Sigma}{\Gamma}{\Judg{\sigma}{B \land C}}}
      {
        \NSeq{\Sigma}{\Gamma}{\Judg{\sigma}{B} }
        \qquad 
        \NSeq{\Sigma}{\Gamma}{\Judg{\sigma}{C}}
      }
$$
$$
\infer[\lorL]
      {\NSeq{\Sigma}
        {\Judg{\sigma}{B \lor C},\Gamma}{\Dscr}}
      {\NSeq{\Sigma}{\Judg{\sigma}{B},\Gamma}{\Dscr}
        \qquad
        \NSeq{\Sigma}{\Judg{\sigma}{C},\Gamma}{\Dscr}
      }
\qquad
\infer[\lorR]
      {\NSeq{\Sigma}{\Gamma}{\Judg{\sigma}{B \lor C}}}
      {\NSeq{\Sigma}{\Gamma}{\Judg{\sigma}{B}}}
$$
$$
\infer[\botL]
      {\NSeq{\Sigma}{\Judg{\sigma}{\bot},\Gamma}{\Bscr}}
      {}
\qquad 
\infer[\lorR]{\NSeq{\Sigma}{\Gamma}{\Judg{\sigma}{B \lor C}}}
        {\NSeq{\Sigma}{\Gamma}{\Judg{\sigma}{C}}}
$$
$$
\infer[\oimpL]{\NSeq{\Sigma}{\Judg{\sigma}{B \oimp C},\Gamma}{\Dscr}}
        {\NSeq{\Sigma}{\Gamma}{\Judg{\sigma}{B}}
        \qquad \NSeq{\Sigma}{\Judg{\sigma}{C},\Gamma}{\Dscr}}
\qquad 
\infer[\oimpR]{\NSeq{\Sigma}{\Gamma}{\Judg{\sigma}{B \oimp C}}}
        {\NSeq{\Sigma}{\Judg{\sigma}{B},\Gamma}{\Judg{\sigma}{C}}}
$$
$$
\infer[\forallL]
      {\NSeq
        {\Sigma}
        {\Judg{\sigma}{\forall_\gamma x.B},\Gamma}
        {\Cscr}
      }
      {
        \TSeq{\Sigma, \sigma}{t}{\gamma}
        \qquad
        \NSeq
        {\Sigma}
        {\Judg{\sigma}{B[t/x]},\Gamma}{\Cscr}
      }
\qquad
\infer[\forallR]
      {\NSeq{\Sigma}{\Gamma}{\Judg{\sigma}{\forall x.B}}}
      {\NSeq{\Sigma, h}{\Gamma}
        {\Judg{\sigma}{B[(h~\sigma)/x]}}}
$$
$$
\infer[\existsL]{\NSeq{\Sigma}{\Judg{\sigma}{\exists x.B},\Gamma}{\Cscr}}
        {\NSeq{\Sigma, h}{\Judg{\sigma}{B[(h~\sigma)/x]},\Gamma}{\Cscr}}
\qquad 
\infer[\existsR]
      {\NSeq{\Sigma}{\Gamma}{\Judg{\sigma}{\exists_\gamma x.B}}}
      {
        \TSeq{\Sigma, \sigma}{t}{\gamma} \qquad
        \NSeq{\Sigma}{\Gamma}{\Judg{\sigma}{B[t/x]}}}
$$
$$
\infer[\nablaL]
{\NSeq{\Sigma}{\Judg{\sigma}{\nabla x\ B},\Gamma}{\Cscr}}
{
\NSeq{\Sigma}{\Judg{(\sigma,y)}{B[y/x]}, \Gamma }{\Cscr}
}
\qquad
\infer[\nablaR]
{\NSeq{\Sigma}{\Gamma}{\Judg{\sigma}{\nabla x\ B}}}
{
\NSeq{\Sigma}{\Gamma}{\Judg{(\sigma,y)}{B[y/x]}}
}
$$
$$
\infer[\cL]{\NSeq{\Sigma}{\Bscr,\Gamma}{\Cscr}}
        {\NSeq{\Sigma}{\Bscr,\Bscr,\Gamma}{\Cscr}}
\qquad
\infer[\wL]
      {\NSeq{\Sigma}{\Bscr, \Gamma}{\Cscr}}
      {\NSeq{\Sigma}{\Gamma}{\Cscr}}
\qquad
\infer[\topR]{\NSeq{\Sigma}{\Gamma}{\Judg{\sigma}{\top}}}{}
$$
}
\caption{The core inference rules of $\FOLNb$.}
\label{fig:folnb}
\end{figure}
\newcommand{\alphaL}{\alpha_{\cal R}}
\newcommand{\alphaR}{\alpha_{\cal L}}

We now show that the formulation of $\LG$ is equivalent 
to $\FOLNb$ extended with the axiom schemes of name permutations
and weakening:
\begin{equation}
\nabla x\nabla y.B \, x\,y \oimp \nabla y\nabla x.B\,x\,y
\quad \hbox{ and } \quad 
B \equiv \nabla x.B
\end{equation}
where $x$ is not free in $B$ in the second scheme.

Sequents in $\FOLNb$ are expressions of the form
$$
\NSeq{\Sigma}{\Judg{\sigma_1}{B_1},\ldots, \Judg{\sigma_n}{B_n}}{\Judg{\sigma_0}{B_0}}.
$$
$\Sigma$ is the {\em signature} of the sequent, $\sigma_i$ is
a list of variables locally scoped over $B_i$, and is referred to as {\em local signature}.
The expression $\Judg{\sigma_i}{B_i}$ is called a {\em local judgment},
or {\em judgment} for short.
In~\cite{miller05tocl}, local judgments are considered equal modulo renaming
of their local signatures, e.g., $(a,b) \triangleright P\,a\,b$ is
equal to $(c,d)\triangleright P\,c\,d.$
Local judgments are ranged over by scripted capital letters, e.g.,
$\Bscr$, $\Dscr$, etc.
For the purpose of proving the correspondence with $\LG$, however,
we will make this renaming step explicit, by including the rules:
$$
\infer[\alphaL, \lambda \vec x.B \equiv_\alpha \lambda \vec y.B']
{\Seq{\Judg{\vec x}{B}, \Gamma}{\Cscr}}
{\Seq{\Judg{\vec y}{B'}, \Gamma}{\Cscr}}
\qquad
\infer[\alphaR,  \lambda \vec x.B \equiv_\alpha \lambda \vec y.B']
{\Seq{\Gamma}{\Judg{\vec x}{B}}}
{\Seq{\Gamma}{\Judg{\vec y}{B'}}}
$$
The inference rules of $\FOLNb$ are given in Figure~\ref{fig:folnb}.

\newcommand{\pL}{p{\cal L}}
\newcommand{\pR}{p{\cal R}}
\newcommand{\ssL}{ss{\cal L}}
\newcommand{\ssR}{ss{\cal R}}
\newcommand{\wsL}{ws{\cal L}}
\newcommand{\wsR}{ws{\cal R}}

We now consider the correspondence between $\LG$ with $\FOLNb$
extended with the following axiom schemes:
\begin{equation}
\label{eq:axiom1}
\nabla x \nabla y.B\,x\,y \equiv \nabla y \nabla x.B\,x\,y.
\end{equation}
\begin{equation}
\label{eq:axiom2}
B \equiv \nabla x.B, \hbox{provided that $x$ is not free in $B$.}
\end{equation}
We can equivalently state these two axioms as the following inference
rules:
$$
\infer[\pL]
{\Seq{\Judg{(\vec x,a,b,\vec y)}{B}, \Gamma}{\Cscr}}
{\Seq{\Judg{(\vec x,b,a,\vec y)}{B}, \Gamma}{\Cscr}}
\qquad
\infer[\pL]
{\Seq{\Gamma}{\Judg{(\vec x,a,b,\vec y)}{B}}}
{\Seq{\Gamma}{\Judg{(\vec x,b,a,\vec y)}{B}}}
$$
$$
\infer[\ssL, a \not \in \{\vec x, \vec y\}]
{\Seq{\Judg{(\vec x\vec y)}{B}, \Gamma}{\Cscr}}
{\Seq{\Judg{(\vec x,a,\vec y)}{B}, \Gamma}{\Cscr}}
\qquad
\infer[\ssR,  a \not \in \{\vec x, \vec y\}]
{\Seq \Gamma {\Judg{(\vec x\vec y)}{B}}}
{\Seq \Gamma {\Judg{(\vec x,a,\vec y)}{B}} }
$$
$$
\infer[\wsL, a \not \in supp(B)]
{\Seq{\Judg{(\vec x,a,\vec y)}{B}, \Gamma}{\Cscr}}
{\Seq{\Judg{(\vec x\vec y)}{B}, \Gamma}{\Cscr}}
\qquad
\infer[\ssR,  a \not \in supp(B)]
{\Seq \Gamma {\Judg{(\vec x,a,\vec y)}{B}}}
{\Seq \Gamma {\Judg{(\vec x\vec y)}{B}} }
$$
Implicit in the above rules is the assumption that
variables in local signatures are considered as
special constants, much like the nominal constants
in $\LG$. The support of $B$, within a local
signature $\sigma$, is defined similarly
as it is in $\LG$: it is the set
$\{ a \in \sigma \mid \hbox{$a$ occurs in $B.$} \}.$

\newcommand{\FOLNbp}{FO\lambda^{\nabla+}}

The logical system with the inference rules in Figure~\ref{fig:folnb} together
with $\alphaL$, $\alphaR$, $\pL$, $\pR$,
$\ssL$, $\ssR$, $\wsL$ and $\wsR$ is referred to as $\FOLNbp$.
In relating $\LG$ and $\FOLNbp$, we map the local signatures
to nominal constants, and vice versa.
In the following, given a formula $B$, we assume a particular
enumeration of the nominal constants appearing in $B$
based the left-to-right order of their appearance in $B$.
\begin{lemma}
\label{lm:lg to folnb}
If the sequent $\NSeq{\Sigma}{B_1,\ldots,B_n}{B_0}$
is provable in $\LG$ then the sequent
$$
\NSeq{\Sigma}{\Judg{\vec c_1}{B_1}, \Judg{\vec c_n}{B_n}}{\Judg{\vec c_0}{B_0}}
$$
where $\vec c_i$  is an enumeration of $supp(B_i),$
is provable in $\FOLNbp.$
\end{lemma}
\begin{proof}
Suppose that $\Pi$ is 
a proof of $\NSeq{\Sigma}{B_1,\ldots,B_n}{B_0}.$
We construct a proof $\Pi'$ of
$$
\NSeq{\Sigma}{\Judg{\vec c_1}{B_1}, \Judg{\vec c_n}{B_n}}{\Judg{\vec c_0}{B_0}}
$$
by induction on $ht(\Pi).$
We consider some interesting cases here:
\begin{itemize}
\item Suppose $\Pi$ ends with  $id_\pi:$
$$
\infer[id_\pi]
{\Seq{\Gamma', B_i}{B_0}}
{\pi.B_i = \pi'.B_0}
$$
The permutations $\pi$ and $\pi'$ can be imitated by a series of renaming
($\alphaL$ and $\alphaR$ rules). 
The derivation $\Pi'$ is therefore constructed by applying a series of 
$\alpha_R$, $\alpha_L$, followed by the $id$ rule.

\item Suppose $\Pi$ ends with $\oimpR:$ in this case we suppose that
$B_0 = C \oimp D.$
$$
\infer[\oimpR]
{\Seq{B_1,\ldots,B_n}{C \oimp D}}
{\deduce{\Seq{B_1,\ldots,B_n, C}{D}}{\Pi_1}}
$$
By induction hypothesis we have a derivation $\Pi_2$
of
$$
\Seq{\Judg{\vec c_1}{B_1}, \ldots, \Judg{\vec c_n}{B_n}, \Judg{\vec a}{C}}
{\Judg{\vec b}{D}}
$$
We first have to weaken the signatures $\vec a$ and $\vec d$ to
$\vec c_0$ before applying the introduction rule for $\oimp$.
That is, $\Pi'$ is the derivation
\settowidth{\infwidthi}
{$\Seq{\Judg{\vec c_1}{B_1,\ldots, \Judg{\vec c_n}{B_n}}, \Judg{\vec c_0}{C}}
    {\Judg{\vec c_0}{D}}$}
$$
\infer[\oimpR]
{\Seq{\Judg{\vec c_1}{B_1,\ldots, \Judg{\vec c_n}{B_n}}}{\Judg{\vec c_0}{C \oimp D}}}
{
\infer[*]
{
\Seq{\Judg{\vec c_1}{B_1,\ldots, \Judg{\vec c_n}{B_n}}, \Judg{\vec c_0}{C}}
    {\Judg{\vec c_0}{D}}
}
{
  \infer[]
   {\makebox[\infwidthi]{}}
   {
        \deduce{\Seq{\Judg{\vec c_1}{B_1,\ldots, \Judg{\vec c_n}{B_n}}, \Judg{\vec a}{C}}
               {\Judg{\vec b}{D}}  }{\Pi_2}
   }
}
}
$$
Here the star `*' denotes a series of applications of $\wsL$,
$\wsR$, $\pL$ and $\pR.$

\item Suppose $\Pi$ is
$$
\infer[\existsR]
{\Seq{B_1,\ldots,B_n}{\exists x.C}}
{\deduce{\Seq{B_1,\ldots,B_n}{C[t/x]}}{\Pi_1}}
$$
It is possible that $t$ contains new constants that are not in the
support of $C.$ Suppose $\vec d$ is an enumeration of 
the support of $C[t/x]$. 
The derivation $\Pi'$ is constructed as follows
$$
\infer[*]
{\Seq{\Judg{\vec c_1}{B_1}, \ldots, \Judg{\vec c_n}{B_n} }
{\Judg{\vec c_0}{\exists x.C}}}
{
\infer[]
{\makebox[\infwidthi]{}}
{
\infer[\existsR]
{ 
\Seq{\Judg{\vec c_1}{B_1}, \ldots, \Judg{\vec c_n}{B_n} }
{\Judg{\vec d}{\exists x.C}}
}
{
\deduce{ 
\Seq{\Judg{\vec c_1}{B_1}, \ldots, \Judg{\vec c_n}{B_n} }
{\Judg{\vec d}{C[t/x]}}
}{\Pi_2}
}
}
}
$$
where $\Pi_2$ is obtained from induction hypothesis applied
to $\Pi_1$, and the rule `*' denotes a series of
applications of $\ssR$ (for introducing new constants) 
and $\pR$ (for rearranging the order of the local signature).

\item For other cases, the construction of $\Pi'$ follows the same 
pattern as in the previous cases, i.e., 
by induction hypothesis, followed by some rearranging, extension,
or weakening of local signatures.
\end{itemize}
\qed
\end{proof}

\begin{lemma}
\label{lm:folnb to lg}
If the sequent 
$$
\NSeq{\Sigma}{\Judg{\vec c_1}{B_1}, \Judg{\vec c_n}{B_n}}{\Judg{\vec c_0}{B_0}}
$$
is provable in $\FOLNbp$ then the sequent
$\NSeq{\Sigma}{B_1,\ldots,B_n}{B_0}$
is provable in $\LG$
\end{lemma}
\begin{proof}
Suppose $\Pi$ is a derivation of 
$$
\Seq{\Judg{\vec c_1}{B_1}, \ldots, \Judg{\vec c_n}{B_n} } {\Judg{\vec c_0}{B_0}}
$$
We construct a derivation $\Pi'$ of 
$\Seq{B_1,\ldots,B_n}{B_0}$ by induction on $ht(\Pi)$.
We show here the interesting cases; the other cases follow
immediately from induction hypothesis:
\begin{itemize}
\item If $\Pi$ ends with $id$, $\topR$, or $\botL$ then $\Pi'$ ends with
the same rule.
\item Suppose $\Pi$ is
$$
\infer[\alphaR]
{\Seq{\Judg{\vec c_1}{B_1}, \ldots, \Judg{\vec c_n}{B_n}}{\Judg{\vec c_0}{B_0}}}
{
\deduce{\Seq{\Judg{\vec c_1}{B_1}, \ldots, \Judg{\vec c_n}{B_n} }{\Judg{\vec d}{B}}}
{\Pi_1}
}
$$
By induction hypothesis, there is a derivation $\Pi_2$
of
$\Seq{{B_1}, \ldots, {B_n} }{{B}}.$
To get $\Pi'$ apply the procedure in Definition~\ref{def:perm}
to $\Pi_2$ to rename $B$ to $B_0$.

\item Suppose $\Pi$ is
$$
\infer[\forallR]
{\Seq{\Judg{\vec c_1}{B_1}, \ldots, \Judg{\vec c_n}{B_n}}{\Judg{\vec c_0}{\forall x.C}}}
{
\deduce{ 
\Seq{\Judg{\vec c_1}{B_1}, \ldots, \Judg{\vec c_n}{B_n}}{\Judg{\vec c_0}{C[(h\,\vec c_0)/x]}}
}{\Pi_1}
}
$$
By induction hypothesis, there is a derivation $\Pi_2$
of 
$\Seq{{B_1}, \ldots, {B_n}}{{C[(h\,\vec c_0)/x]}}.$
Suppose $\{ \vec d \} = supp(C).$ Then $\Pi'$ is
$$
\infer[\forallR]
{\Seq{B_1,\ldots,B_n}{\forall x.C}}
{
\deduce{\Seq{B_1,\ldots,B_n}{C[h'\,\vec d/x]}}{\Pi_2[\lambda \vec c_0.h'\,\vec d/h]}
}
$$

\item If $\Pi$ ends with $\existsL$, apply the same construction as in the
previous case.
\end{itemize}
\qed
\end{proof}

\begin{theorem}
\label{thm:lg equal folnb}
Let $F$ be a formula which contains no occurrences of 
nominal constants. Then $F$ is provable in $\FOLNb$ extended with
the axiom schemes $B \equiv \nabla x.B$ and 
$\nabla x\nabla y. B\,x\,y \oimp \nabla y\nabla x.B\,x\,y$
if and only if 
$F$ is provable in $\LG.$
\end{theorem}


\begin{thebibliography}{10}

\bibitem{girard92mail}
J.-Y. Girard.
\newblock A fixpoint theorem in linear logic.
\newblock Email to the linear@cs.stanford.edu mailing list, February 1992.

\bibitem{hallnas91jlc}
L.~Halln{\"{a}}s and P.~Schroeder-Heister.
\newblock A proof-theoretic approach to logic programming. {II}. {Programs} as
  definitions.
\newblock {\em Journal of Logic and Computation}, 1(5):635--660, October 1991.

\bibitem{mcdowell00tcs}
R.~McDowell and D.~Miller.
\newblock Cut-elimination for a logic with definitions and induction.
\newblock {\em Theoretical Computer Science}, 232:91--119, 2000.

\bibitem{mcdowell02tocl}
R.~McDowell and D.~Miller.
\newblock Reasoning with higher-order abstract syntax in a logical framework.
\newblock {\em ACM Transactions on Computational Logic}, 3(1):80--136, January
  2002.

\bibitem{miller91jlc}
D.~Miller.
\newblock A logic programming language with lambda-abstraction, function
  variables, and simple unification.
\newblock {\em Journal of Logic and Computation}, 1(4):497--536, 1991.

\bibitem{miller92jsc}
D.~Miller.
\newblock Unification under a mixed prefix.
\newblock {\em Journal of Symbolic Computation}, 14(4):321--358, 1992.

\bibitem{miller99surveys}
D.~Miller and C.~Palamidessi.
\newblock Foundational aspects of syntax.
\newblock In P.~Degano, R.~Gorrieri, A.~Marchetti-Spaccamela, and P.~Wegner,
  editors, {\em ACM Computing Surveys Symposium on Theoretical Computer
  Science: A Perspective}, volume~31. ACM, September 1999.

\bibitem{miller05tocl}
D.~Miller and A.~Tiu.
\newblock A proof theory for generic judgments.
\newblock {\em ACM Trans.\ on Computational Logic}, 6(4):749--783, Oct. 2005.

\bibitem{nipkow93lics}
T.~Nipkow.
\newblock Functional unification of higher-order patterns.
\newblock In M.~Vardi, editor, {\em Proc.\ 8th {IEEE} Symposium on Logic in
  Computer Science ({LICS} 1993)}, pages 64--74. IEEE, June 1993.

\bibitem{pfenning88pldi}
F.~Pfenning and C.~Elliott.
\newblock Higher-order abstract syntax.
\newblock In {\em Proceedings of the {ACM}-{SIGPLAN} Conference on Programming
  Language Design and Implementation}, pages 199--208. ACM Press, June 1988.

\bibitem{pitts03ic}
A.~M. Pitts.
\newblock Nominal logic, a first order theory of names and binding.
\newblock {\em Information and Computation}, 186(2):165--193, 2003.

\bibitem{schroeder-heister92nlip}
P.~Schroeder-Heister.
\newblock Cut-elimination in logics with definitional reflection.
\newblock In D.~Pearce and H.~Wansing, editors, {\em Nonclassical Logics and
  Information Processing}, volume 619 of {\em LNCS}, pages 146--171. Springer,
  1992.

\bibitem{tiu04phd}
A.~Tiu.
\newblock {\em A Logical Framework for Reasoning about Logical Specifications}.
\newblock PhD thesis, Pennsylvania State University, May 2004.

\bibitem{tiu07entcs}
A.~Tiu.
\newblock A logic for reasoning about generic judgments.
\newblock {\em Electr. Notes Theor. Comput. Sci.}, 174(5):3--18, 2007.

\end{thebibliography}
\end{document}